\documentclass[a4paper,preprint,nofootinbib]{revtex4}
\usepackage{graphicx}
\usepackage{amsmath}
\usepackage{amssymb}
\usepackage{float}
\usepackage{natbib}

\newcommand{\Slash}[1]{{\ooalign{\hfil/\hfil\crcr$#1$}}} 

\newcommand{\nn}{\nonumber}
\newcommand{\be}{\begin{eqnarray}}
\newcommand{\ee}{\end{eqnarray}}
\catcode`\@=11

\def\lsim{\mathrel{\mathpalette\@versim<}}
\def\gsim{\mathrel{\mathpalette\@versim>}}
\def\@versim#1#2{\vcenter{\offinterlineskip
\ialign{$\m@th#1\hfil##\hfil$\crcr#2\crcr\sim\crcr } }}
\catcode`\@=12

\begin{document}

\title{Gravitational Waves from Hidden QCD Phase Transition}

\author{Mayumi \surname{Aoki}}
\email{mayumi@hep.s.kanazawa-u.ac.jp}
\affiliation{Institute for Theoretical Physics, Kanazawa University, Kanazawa 920-1192, Japan}

\author{Hiromitsu \surname{Goto}}
\email{goto@hep.s.kanazawa-u.ac.jp}
\affiliation{Institute for Theoretical Physics, Kanazawa University, Kanazawa 920-1192, Japan}

\author{Jisuke \surname{Kubo}}
\email{jik@hep.s.kanazawa-u.ac.jp}
\affiliation{Institute for Theoretical Physics, Kanazawa University, Kanazawa 920-1192, Japan}

\preprint{KANAZAWA-17-08}

%=======================================================================%
%
\begin{abstract} 
Drastic changes in the early Universe such as first-order phase transition can produce a stochastic gravitational wave (GW) background.
We investigate  the testability of a scale invariant extension of the standard model (SM) using the GW background produced 
by the chiral phase transition in a strongly interacting QCD-like hidden sector,
which, via a SM singlet real scalar mediator, triggers the electroweak phase transition. 
Using the Nambu--Jona-Lasinio method in a mean field approximation
we estimate the GW signal and
find that it can be tested by future space-based detectors.
\end{abstract}
\maketitle

%=======================================================================%
%=======================================================================%
%
\section{Introduction}
It is a challenge for physics beyond the standard model (SM) 
to answer a long-standing question---what is the origin of mass?
The same question applies to dark matter (DM),
which, if it is a particle, is absent in the SM.
Though various suggestions about how to go beyond the SM exist,
there is so far no sign for that 
from the Large Hadron Collider (LHC)  experiments  \cite{ATLAS,CMS}
and no sign from the current DM detection experiments either \cite{Akerib:2016vxi,Aprile:2017iyp}.

In contrast to this situation, 
the first observation of the gravitational wave (GW) signal at LIGO \cite{Abbott:2016blz} has opened up a new way to study astrophysical phenomena
and has awakened the hope in particle cosmology that 
phenomena in the early Universe can also be probed by the GW.
It has indeed been known that phenomena in the early Universe
such as inflation \cite{Starobinsky:1979ty}, topological defects \cite{Vilenkin:2000jqa}, and first-order phase transition \cite{Witten:1984rs} generate a nonlocalized stochastic GW background.
In particular, phase transitions in particle physics are associated with symmetry breaking,
and therefore the GW signals produced by these phase transitions can be an alternative approach 
to investigate the structure of symmetries
in the early Universe.
Unfortunately,  because of not being first order,
the phase transition associated with the electroweak (EW) symmetry breaking in the SM 
cannot produce the GW background \cite{Kajantie:1995kf,Kajantie:1996mn,Rummukainen:1998as}.
However, if the SM is extended, 
observable GW signals associated with a symmetry breaking
may be produced and tested in future experiments 
such as LISA \cite{Caprini:2015zlo,Audley:2017drz} and DECIGO \cite{Seto:2001qf,Kawamura:2006up,Kawamura:2011zz}
as discussed in \cite{Apreda:2001us,Grojean:2006bp,Espinosa:2008kw,Ashoorioon:2009nf,Das:2009ue,Schwaller:2015tja,Kakizaki:2015wua,Jinno:2016knw,Hashino:2016rvx,Kubo:2016kpb,Vaskonen:2016yiu,Beniwal:2017eik,Marzola:2017jzl,Chao:2017vrq,Bian:2017wfv,Chao:2017ilw,Dev:2016feu}.

As lattice simulations in QCD have shown
\cite{Aoki:2006we,Petreczky:2012rq,Bhattacharya:2014ara},
the chiral phase transition in QCD is,
due to a relatively large current  mass of the strange quark, a crossover type.
This does not prevent the possibility
that the chiral phase transition in a QCD-like hidden sector is of first order.\footnote{
Other possibilities of a first order phase transition in a QCD-like theory are the deconfinment/confinement phase transition in the quenched QCD \cite{Petreczky:2012rq} and the chiral phase transition in QCD with a large baryon chemical potential \cite{Fodor:2004nz,Schwarz:2009ii}. They may produce the observable GW signal as discussed in \cite{Ahmadvand:2017xrw,Caprini:2010xv,Ahmadvand:2017tue}.}
In fact, such a  possibility with a critical 
temperature of $\mathcal{O}(1)$ TeV has been recently 
found \cite{Holthausen:2013ota,Ametani:2015jla} in  a scale invariant extension of the SM
\cite{Hur:2007uz,Hur:2011sv,Holthausen:2013ota,Heikinheimo:2013fta,Kubo:2014ida,Ametani:2015jla,Hatanaka:2016rek}, in which dynamical chiral symmetry breaking  (D$\chi$SB) in a QCD-like hidden sector triggers the EW symmetry breaking.
In the present paper we focus on this model.
In this model, moreover, the EW energy scale and the DM mass have the same origin.
In most of the parameter space, the DM mass is created before
the EW phase transition and,
in a certain region of the parameter space,
it takes place during a strong first-order chiral phase transition.
By choosing various benchmark points in the parameter space
we study the testability of the GW background produced by this
phase transition.

The paper is organized as follows.
In Sec. \ref{Hidden QCD Physics} we briefly review the scale invariant extension of the SM with a QCD-like hidden sector and 
describe how we use the Nambu--Jona-Lasinio (NJL) model 
\cite{Nambu:1960xd,Nambu:1961tp,Nambu:1961fr}
as an effective low-energy theory in a mean field approximation
\cite{Kunihiro:1983ej,Hatsuda:1994pi}. 
We fix the number of the hidden color $n_c$ and flavor $n_f$
both at $3$, because we can simply  rescale the values of the NJL parameters for the real hadrons.
In this way we can avoid increasing the number of independent
parameters when going from the high-energy theory to the low-energy effective theory.
We pick up a set of four benchmark parameters, for which
the chiral phase transition in the hidden QCD sector  is of first order.
For these points we calculate 
the GW signals.

Note that the chiral phase transition in our model
occurs in a two-dimensional parameter space, the chiral condensate
and the vacuum expectation value (VEV) of the SM singlet real scalar (which is the mediator
of the energy scale from the hidden sector to the SM sector).
Furthermore, the mean field $\sigma$ corresponding to
the chiral condensate is a nonpropagating field at the zeroth order
in the mean field approximation: It becomes a quantum field 
at the one-loop level, so that its wave function renormalization
constant is far from $1$ and depends on the mean fields as well as on the
temperature.
In Sec.~\ref{Bubbles from the Tunneling
in the Hidden Sector} we discuss how to
manage the complications mentioned above to
compute the rate of  the bubble nucleation that occurs during
the cosmological tunneling in the   hidden QCD sector.
In Sec.~\ref{Signal From Hidden QCD} we discuss the detectability of the GW signals 
produced by the chiral phase transition 
for the  benchmark points in the parameter  space.
We summarize and conclude  in Sec.~\ref{Summary and Conclusion}.

%=======================================================================%
%=======================================================================%
\section{The Model}\label{Hidden QCD Physics}

We consider a classically scale invariant extension of the SM studied in 
\cite{Hur:2007uz,Hur:2011sv,Holthausen:2013ota,Heikinheimo:2013fta,Ametani:2015jla}, which consists of a hidden $\mathrm{SU}(n_c)_{\mathrm{H}}$ gauge sector coupled via a real singlet scalar field $S$ to the SM.
The Lagrangian of the hidden sector is written as 
\begin{align}
\label{LH}
\mathcal{L}_{\mathrm{H}}=-\frac{1}{2}\mathrm{Tr}~F^2 + \mathrm{Tr}~ \bar{\psi}\left( i\Slash{D}-yS\right)\psi,
\end{align}
where the hidden vectorlike fermions $\psi_i ~(i=1,\dots,n_f)$ transform as a fundamental representation of $\mathrm{SU}(n_c)_{\mathrm{H}}$.
The $\mathcal{L}_{\text{SM}+S}$ part of the total Lagrangian $\mathcal{L}_{\mathrm{T}}=\mathcal{L}_{\text{SM}+S}+\mathcal{L}_{\text{H}}$ contains the SM gauge and Yukawa interactions along with the scalar potential
\begin{align}
\label{VSMS}
V_{\mathrm{SM}+S}=\lambda_H (H^{\dag}H)^2+\frac{1}{4}\lambda_S S^4-\frac{1}{2}\lambda_{HS}S^2(H^{\dag}H),
\end{align}
where $H^T=(H^+,(h+G)/\sqrt{2})$ is the SM Higgs doublet field with $H^+$ and $G$ as the would-be Nambu-Goldstone (NG) fields.
The scalar couplings at the tree level
have to satisfy the stability condition for the scalar potential
\begin{align}
\label{stability}
    \lambda_{H}>0,~~\lambda_{S}>0~~~\mathrm{and}~~~2\sqrt{\lambda_{H}\lambda_{S}}-\lambda_{HS}&>0.
\end{align}
Here $y$ and $\lambda_{HS}$ are assumed to be positive.
This model explains the origin of the mass of the Higgs boson and the DM in the following
sense.
\begin{itemize}
  \item First, due to the D$\chi$SB in the hidden sector,  a nonzero  chiral 
  condensate $\left<\bar{\psi}\psi\right>$  forms and
generates a mass scale above the EW scale. 
  Consequently, NG bosons, which are  mesons
  in the hidden sector, appear.
  \item At the same time of the hidden D$\chi$SB, the singlet scalar field $S$ acquires a nonzero VEV $\left<S\right>$ because of the Yukawa interaction $-yS\bar{\psi}\psi$.
Note that the Yukawa interaction breaks the chiral symmetry explicitly, and $y\left<S\right>$ plays the role of a current mass. Therefore,
  the mass of the hidden mesons depends crucially on $y\left<S\right>$.
  \item These hidden mesons (or a part of them) can become DM candidates, because they are stable due to the vectorlike flavor symmetry that is left unbroken after the 
  D$\chi$SB.
  \item The EW symmetry breaking is triggered by the Higgs mass term that is nothing 
  but the scalar coupling $+\frac{1}{2}\lambda_{HS}S^2H^{\dag}H$ with the nonzero $\left< S \right>$.
  \end{itemize}

In this work we consider the case with $n_c=n_f=3$ and assume that
 the singlet scalar $S$ equally couples to the hidden fermions.
Then the hidden chiral symmetry $\mathrm{SU}(3)_{L}\times\mathrm{SU}(3)_{R}$ is dynamically broken  down to $\mathrm{SU}(3)_{V}$, and 
thanks to this unbroken  symmetry, eight hidden pions become a DM candidate.
The DM physics and the impact of the hidden chiral phase transition to the EW phase transition
have been investigated in \cite{Holthausen:2013ota} by using the NJL theory \cite{Nambu:1960xd,Nambu:1961tp,Nambu:1961fr} in the self-consistent mean field (SCMF) approximation \cite{Hatsuda:1994pi,Kunihiro:1983ej}.
It has been found that a strong first-order chiral phase transition can occur if the Yukawa coupling $y$ is small enough, i.e., $y\lesssim 0.006$ \cite{Ametani:2015jla}.
Within the framework of  the NJL theory we will calculate the GW spectrum 
produced by the hidden chiral phase transition later on.
The same model has been analyzed
 by using a linear \cite{Hur:2007uz} and nonlinear \cite{Hur:2011sv} sigma model and also  AdS/QCD approach \cite{Hatanaka:2016rek}.
In \cite{Tsumura:2017knk},   the GW spectrum from the hidden chiral phase transition has been calculated within the framework of a linear sigma model.

%=======================================================================%
\subsection{Nambu--Jona-Lasinio Lagrangian in a Mean Field Approximation}
Following \cite{Holthausen:2013ota} we approximate the high-energy Lagrangian (\ref{LH}) by
the NJL Lagrangian
\begin{align}
\label{LNJL}
\mathcal{L}_{\mathrm{NJL}}= \mathrm{Tr} ~\bar{\psi}\left(i\Slash{\partial}-yS\right)\psi+2G\mathrm{Tr}~\Phi^{\dag}\Phi +G_D(\mathrm{det}~\Phi+h.c. ),
\end{align}
where $G$ and $G_D$ are dimensional parameters and
\begin{align}
  (\Phi)_{ij}  & = \bar{\psi}_i (1-\gamma_5)\psi_j =\frac{1}{2}\lambda^a _{ji}\mathrm{Tr}~ \bar{\psi}\lambda^a (1-\gamma_5)\psi,  \nn\\
 (\Phi^{\dag})_{ij}  & = \bar{\psi}_i (1+\gamma_5)\psi_j =\frac{1}{2}\lambda^a _{ji}\mathrm{Tr} ~\bar{\psi}\lambda^a (1+\gamma_5)\psi.\nn
\end{align}
Here $\lambda^a~(a=1,\cdots,8)$ are the Gell-Mann matrices with $\lambda^0=\sqrt{2/3}~\bm{1}$.
To deal with the NJL Lagrangian (\ref{LNJL}), 
which is nonrenormalizable,
we work in the SCMF approximation 
\cite{Hatsuda:1994pi,Kunihiro:1983ej}.
The mean fields $\sigma$ and $\phi_a$ are defined in the Bardeen-Cooper-Schrieffer vacuum as 
\begin{align}
\label{varphi}
\left<\Phi\right>  = -\frac{1}{4G}\left(\text{diag}(\sigma, \sigma, \sigma) +i\left( \lambda^a\right)^T \phi_a \right).
\end{align}
After splitting up the NJL Lagrangian into the sum $\mathcal{L}_{\text{NJL}} =\mathcal{L}_{\text{MFA}}+\mathcal{L}_{I}$,
where $\mathcal{L}_{\text{MFA}}$ contains at most bilinear terms 
of $\psi_i$ and  $\mathcal{L}_{I}$ is normal ordered with respect to  the BCS vacuum,
we find the Lagrangian in the SCMF approximation:
\begin{align}
\nonumber
   \mathcal{L}_{\text{MFA}} = &   \mathrm{Tr} ~\bar{\psi}(i\Slash{\partial}-M)\psi -i\mathrm{Tr} ~\bar{\psi}\gamma_5 \phi \psi -\frac{1}{8G}\left( 3\sigma^2+2\sum^8 _{a=1} \phi_a \phi_a \right)  \\
 \label{Hidden SCMFA}
    & +\frac{G_D}{8G^2}\left(  -\mathrm{Tr} ~\bar{\psi} \phi^2 \psi + \sum^8 _{a=1} \phi_a \phi_a \mathrm{Tr} ~\bar{\psi}\psi + i\sigma \mathrm{Tr}~ \bar{\psi}\gamma_5 \phi \psi +\frac{\sigma^3}{2G}+\frac{\sigma}{2G}\sum^8 _{a=1} (\phi_a)^2   \right) ,
\end{align}
where $\phi=\sum_{a=1}^8~\phi_a \lambda^a$, we have suppressed $\phi_0$ here, and $M$ is given by 
\begin{align}
\label{M}
M= \sigma+yS-\frac{G_D}{8G^2}\sigma^2.
\end{align}
Through integrating out the hidden fermions, a nontrivial correction to the tree-level potential for $\sigma$ is generated, 
such that the position of the potential minimum can be shifted from zero to a finite value of $\sigma$.
From the definition (\ref{varphi}) we see that this is nothing but the chiral condensate
 in the SCMF approximation.
By self-consistency it is  meant that the actual value of  
$\langle \sigma\rangle$ is computed afterward at the loop level, and then we consider the mean field Lagrangian
(\ref{Hidden SCMFA}) around this mean field vacuum.
At the tree level of (\ref{Hidden SCMFA}), the mean fields $\sigma$ and $\phi_a$ are nonpropagating classical fields.
Through  integrating out the hidden fermions at the one-loop level, 
their kinetic terms are also generated.
At this stage we reinterpret them as propagating quantum fields.

%=======================================================================%
\subsection{Mass Spectrum}
The chiral condensation in the hidden sector can be studied by using the one-loop effective potential obtained from the mean field Lagrangian (\ref{Hidden SCMFA}):
\be
V_{\mathrm{eff}}= V_{\mathrm{SM}+S}+V_{\mathrm{NJL}},
\ee
where
\begin{align}
   V_{\mathrm{NJL}} (\sigma,S;\Lambda_{\mathrm{H}}) = \frac{3}{8G}\sigma^2-\frac{G_D}{16G^3}\sigma^3-3n_cI_0(M;\Lambda_{\mathrm{H}}),
   \label{Potential0T}
\end{align}
and $I_0$ is given by
\begin{align}
  I_0(M;\Lambda)= \frac{1}{16\pi^2}\left[ \Lambda^4 \ln \left( 1+\frac{M^2}{\Lambda^2 }\right)-M^4 \ln \left( 1+\frac{\Lambda^2 }{M ^2}\right) + \Lambda^2 M ^2\right] .
\end{align}
Here we have used the four-dimensional cutoff,
and $\Lambda$ is the corresponding  cutoff parameter.
The NJL parameters for the hidden QCD are obtained by scaling up the values of
$G, G_D$, and $\Lambda$ for the real hadrons. 
That is, we assume that the dimensionless combinations
\begin{align}
 G^{1/2}\Lambda=1.82,~~~~(-G_D)^{1/5}\Lambda=2.29,
 \label{NJL para}
\end{align}
which are satisfied for the real hadrons, remain unchanged for a higher scale of $\Lambda$.
For a given set of the free parameters of the model $\lambda_H, \lambda_{HS}, \lambda_{S}$, and $y$,
the VEV of $\sigma$ and $S$ can be determined through the minimization of the scalar potential $V_{\mathrm{eff}}(h,S,\sigma;\Lambda_{\mathrm{H}})$,
where the hidden QCD scale $\Lambda_{\mathrm{H}}$ is so chosen to satisfy $\left<h\right>=246~\mathrm{GeV}$.

The mass spectrum of the particles can be computed from the corresponding two-point functions, which 
are obtained by integrating out the hidden fermions.
Note that the $CP$-even scalars $h, S$, and $\sigma$ mix with one another.
The flavor eigenstates $\varphi_i ~(i=h, S, \sigma)$ and the mass eigenstates $s_i ~(i=1, 2, 3)$
are related by $\varphi_i=\xi_{i}^{(j)}s_j $.
Their masses are determined by the zeros of the two-point functions $\Gamma_{ij}~(i,j=h,S,\sigma)$ at the one-loop level,
i.e., $\Gamma_{ij}(m^2_k)\xi_{j}^{(k)}=0$, where
\begin{align}
\Gamma_{hh}(p^2)&=p^2-3\lambda_{H}\left<h\right>^2+\frac{1}{2}\lambda_{HS}\left<S\right>^2,~~~~\Gamma_{hS}=\lambda_{HS}\left<h\right>\left<S\right>,~~~~\Gamma_{h\sigma}=0,  \nonumber \\
\Gamma_{SS}(p^2)&=p^2-3\lambda_{S}\left<S\right>^2+\frac{1}{2}\lambda_{HS}\left<h\right>^2-y^23n_cI_{\varphi^2}(p^2,M;\Lambda_{\mathrm{H}}), \nonumber \\
\Gamma_{S\sigma}(p^2)&=-y\left(1-\frac{G_D\left<\sigma\right>}{4G^2}\right)3n_cI_{\varphi^2}(p^2,M;\Lambda_{\mathrm{H}}), 
\nn\\
\Gamma_{\sigma\sigma}(p^2)&=-\frac{3}{4G}+\frac{3G_D\left<\sigma\right>}{8G^3}-\left(1-\frac{G_D\left<\sigma\right>}{4G^2}\right)^23n_cI_{\varphi^2}(p^2,M;\Lambda_{\mathrm{H}}) 
\label{CPEVENscalar}\\
 &
+\frac{G_D}{G^2}3n_cI_V(M;\Lambda_{\mathrm{H}}),\nn
\end{align}
and the loop functions are defined as
\begin{align}
 \label{propagator for sigma}
  I_{\varphi^2}(p^2,M;\Lambda) &= \int_{\Lambda} \frac{d^4 k}{i(2\pi)^4}\frac{\mathrm{Tr}(\Slash{k}+\Slash{p}+M)(\Slash{k}+M)}{((k+p)^2-M^2)(k^2-M^2)}, \\
  I_{V}(M;\Lambda) & = \int_{\Lambda}  \frac{d^4 k}{i(2\pi)^4}\frac{M}{(k^2-M^2)}=-\frac{1}{16\pi^2}M\left[ \Lambda^2-M^2\ln \left( 1+\frac{\Lambda^2}{M^2}\right) \right].
\end{align}
We identify the SM Higgs with the mass eigenstate corresponding to
 $\xi_{1}$, which is supposed to be closest to $(1,0,0)$,
and its mass is  $m_1=m_h=125.09\pm 0.24~\mathrm{GeV}$ \cite{Olive:2016xmw}.
Similarly we use
$m_2=m_S$ and $m_3=m_{\sigma}$.
The DM candidate is the hidden pion $\phi_a$ and its mass is also generated at the one-loop level.
Its two-point function is 
\begin{align}
\label{GammaDM}
\Gamma_{\mathrm{DM}}(p^2)&=-\frac{1}{2G}+\frac{G_D\left<\sigma\right>}{8G^3}+\left(1-\frac{G_D\left<\sigma\right>}{8G^2}\right)^22n_cI_{\phi^2}(p^2,M;\Lambda_{\mathrm{H}}) +\frac{G_D}{G^2}n_cI_V(M;\Lambda_{\mathrm{H}}),
\end{align}
where the loop function is given by
\begin{align}
  I_{\phi^2}(p^2,M;\Lambda) &= \int_{\Lambda}  \frac{d^4 k}{i(2\pi)^4}\frac{\mathrm{Tr}(\Slash{k}-\Slash{p}+M)\gamma_5(\Slash{k}+M)\gamma_5}{((k-p)^2-M^2)(k^2-M^2)}.
\end{align}
Then we can calculate the DM mass from  $\Gamma_{\mathrm{DM}}(m_{\mathrm{DM}}^2)=0$.

\begin{figure}[t]
\begin{minipage}{0.5\hsize}
\begin{center}
\includegraphics[width=3.1in]{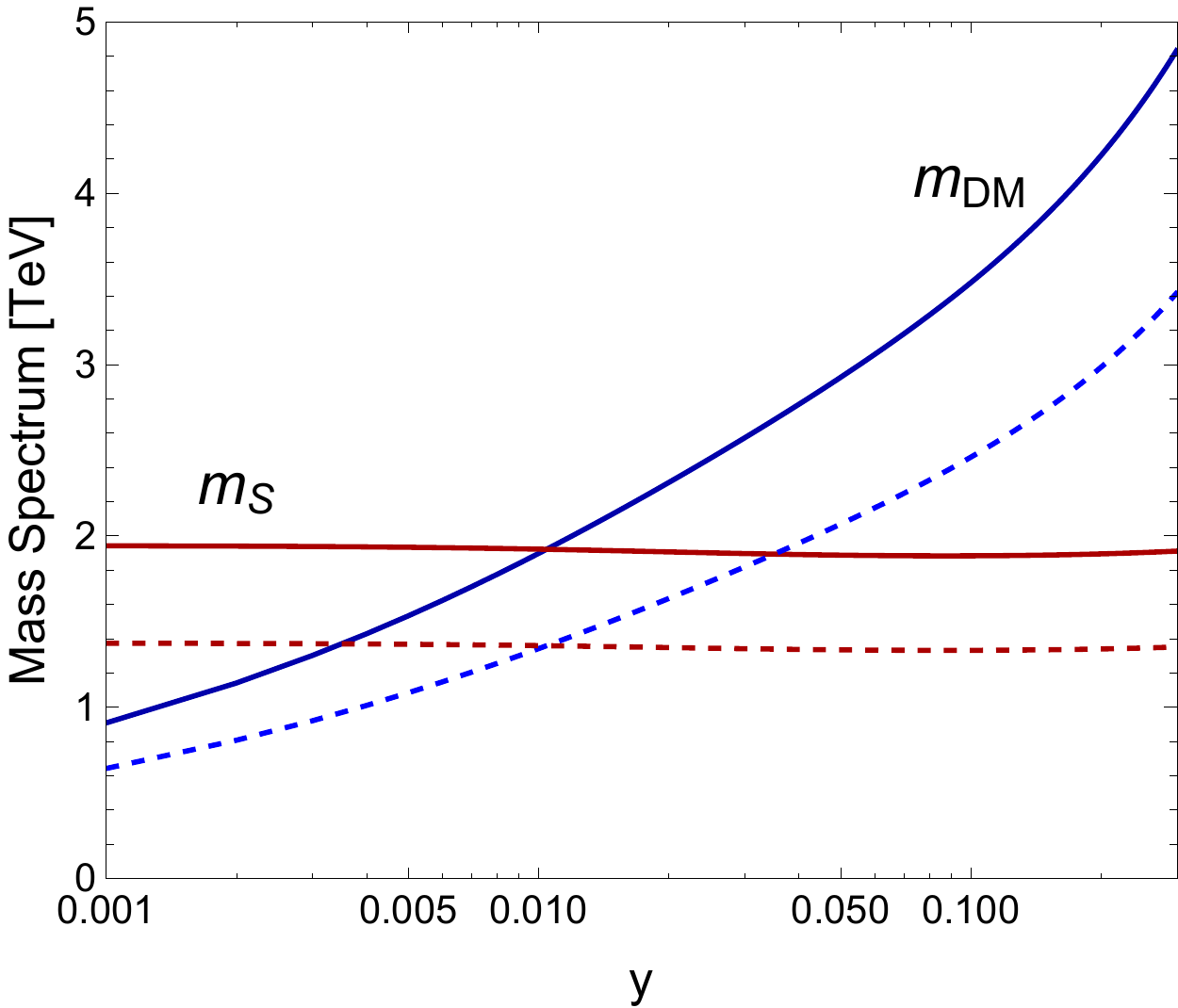}
\end{center}
\end{minipage}
\begin{minipage}{0.49\hsize}
\begin{center}
\includegraphics[width=2.8in]{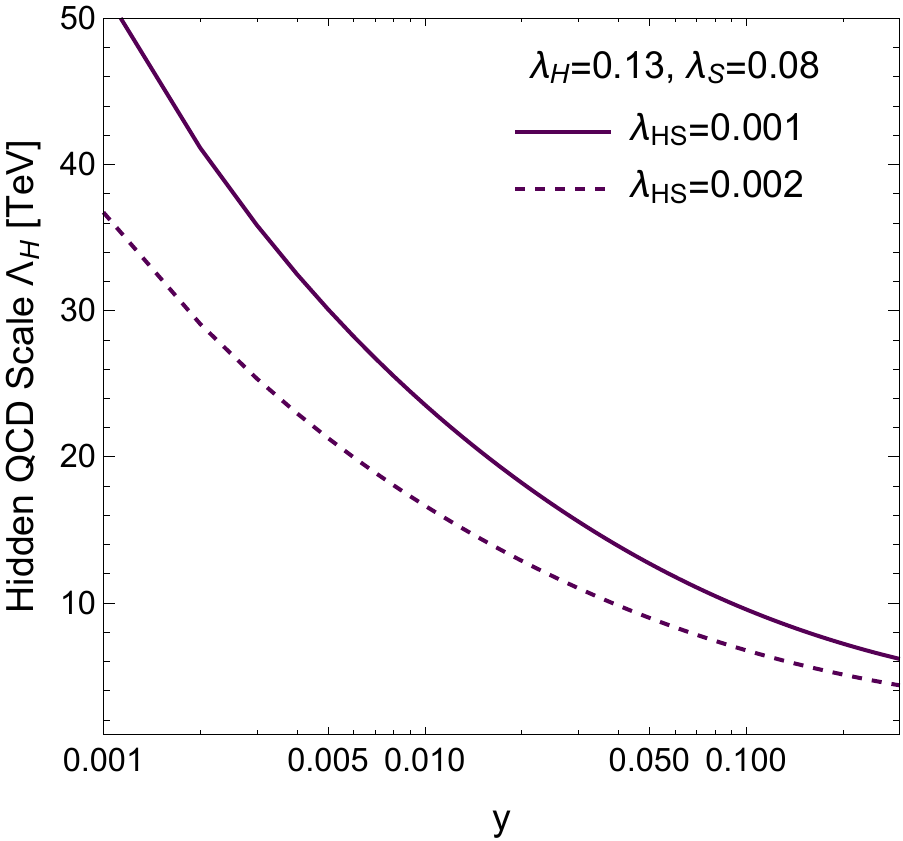}
\end{center}
\end{minipage}
\caption{The masses $(m_{\mathrm{DM}}, m_S)$ (left) and the hidden QCD scale $\Lambda_{\mathrm{H}}$ (right) versus $y$ for $\lambda_{H}=0.13$, $\lambda_{S}=0.08$ with two different values of $\lambda_{HS}$; $\lambda_{HS}=0.001$ (solid lines) and $0.002$ (dashed lines).}
\label{MassSpectrum}
\end{figure}

Once the set of the parameters $(\lambda_{H},\lambda_{HS},\lambda_{S},y)$ is given, the mass spectrum of the hidden sector particles is fixed.
Figure~\ref{MassSpectrum} shows the Yukawa coupling $y$ dependence of the masses $(m_{\mathrm{DM}},m_S)$ (left) and of the hidden QCD scale $\Lambda_{\mathrm{H}}$ (right) for $\lambda_{H}=0.13$, $\lambda_{S}=0.08$ with two different values of $\lambda_{HS}$;  $\lambda_{HS}=0.001$ (solid lines) and $0.002$ (dashed lines).
As shown in Fig.~\ref{MassSpectrum} (left),
the DM mass $m_{\mathrm{DM}}$ is proportional to the Yukawa coupling $y$.
This is because  the Yukawa interaction breaks the chiral symmetry explicitly.
The scale of the D$\chi$SB in the hidden sector, which is the hidden QCD scale $\Lambda_{\mathrm{H}}$,
depends on how the mediator $S$ transfers the mass scale to the SM sector.
The larger the couplings $\lambda_{HS}$ and $y$ are, 
the closer to the EW scale the hidden QCD scale $\Lambda_{\mathrm{H}}$ is located as seen in Fig.~\ref{MassSpectrum} (right). 
Moreover the annihilation processes of the DM also depend on the mass spectrum and the Yukawa coupling $y$.
Note that the one-loop effective couplings are given by $\Gamma_{\phi\phi S}\propto y$ and $\Gamma_{\phi\phi SS}\propto y^2$.
In the small $y$ area with $m_S>m_{\mathrm{DM}}$, 
the mass spectrum should satisfy the resonance condition $m_S\simeq2m_{\mathrm{DM}}$ to obtain a realistic DM relic abundance
and, in this parameter space, the spin-independent  cross section of DM off
the nucleon becomes so small \cite{Holthausen:2013ota}
that it will be very difficult to detect
 DM at direct DM detection experiments such as XENON1T \cite{Aprile:2017iyp}.
On the other hand,
the GW signal might be observed since a strong first-order chiral phase transition can appear for a small $y$ area \cite{Ametani:2015jla}.

%=============   \ref{MassSpectrum}   ==============%
%=================================%

%=======================================================================%
\subsection{Chiral Phase Transitions}\label{hQCD-PT}

The phase transition at finite temperature can be studied using a one-loop effective potential.
Since the EW phase transition occurs well below the critical temperature of the chiral phase transition 
in the hidden sector, we may assume $\left<h\right>=0$ in investigating the chiral phase transition.
Accordingly, the scalar potential to be analyzed is
\begin{align}
\label{VEFF}
   V_{\text{EFF}} (S,\sigma,T) =  V^{h\rightarrow0}_{\text{SM}+S}(S)+V_{\text{NJL}} (S,\sigma)+V_{\text{CW}} (S)+V_{\text{FT}}(S,\sigma,T)+V_{\text{RING}}(S,T) ,
\end{align}
where $V_{\text{SM}+S}$ and $V_{\text{NJL}}$ are given, respectively, in (\ref{VSMS}) and (\ref{Potential0T}), 
\begin{align}
V_{\text{CW}} (S)&= -\frac{9}{4}\frac{\lambda^2 _S}{32\pi^2}(S^4 -\left<S\right>^4)+\frac{m^4_S(S)}{64\pi^2}\ln \left[ \frac{m^2_{S}(S)}{m^2_S(\left<S\right>)}\right],\\
V_{\text{FT}}(S,\sigma,T)&= \frac{T^4}{2\pi^2}J_{B}(m_S ^2(S)/T^2) -6n_c\frac{T^4}{\pi^2}J_{F}(M^2(S,\sigma)/T^2),\\
V_{\text{RING}}(S,T)&= -\frac{T}{12\pi}\left[(M^2 _S(S,T))^{3/2}-(m_S ^2(S))^{3/2}\right],
\end{align}
and  $m_S^2(S)=3\lambda_SS^2+\mathcal{O}(y^2)$ 
 is the field-dependent mass for $S$ with
its thermal mass 
 \begin{align}
M^2 _S = m_S ^2(S) +\left(\frac{\lambda_S}{4}-\frac{\lambda_{HS}}{6}\right)T^2.
\end{align}
The thermal function is
\begin{align}
\label{thermalFB}
  J_{B,F}(r^2) & =\int^{\infty} _{0} dx x^2 \ln \left( 1\mp e^{-\sqrt{x^2+r^2}}\right),
\end{align}
for which we use the approximate expression
\begin{align}
 J_{B,F}(r^2) & =e^{-r^2}\sum^{40} _{n=0} c^{B,F}_n r^{2n}.
 \end{align}

%=======================================================================%
\subsection{Benchmark Points}
As discussed in \cite{Ametani:2015jla},
the chiral phase transition in the hidden sector becomes first order for small $y\lesssim 0.006$.
We require the perturbativity and stability condition (\ref{stability}) of the scalar potential for $y\lesssim 0.006$ to be satisfied
up to the Planck scale at the one-loop level.\footnote{
According to \cite{Bardeen:1995kv},
the hierarchy problem  can be 
 avoided  in this way at least at the one-loop level.} We find that 
\begin{align}
\label{scalars}
  0.13\lesssim\lambda_{H}\lesssim 0.14,  &  ~~~0<\lambda_{HS}< 0.12,  ~~~4\lambda_{HS}^2/\lambda_H<\lambda_{S}\lesssim 0.23
\end{align}
should be satisfied to meet the requirements.
The inequality $0<\lambda_{HS}$ is our assumption (see (\ref{stability})), 
and 
the interval of $\lambda_{H}$ is due to the observed Higgs mass.
The upper limit of $\lambda_{S}$ comes from perturbativity,
while the lower limit comes from the stability condition with finite $\lambda_{H}$ and $\lambda_{HS}$.
Note that there is no lower limit on $\lambda_{HS}$ and $y$.
We however consider only the case for $\lambda_{HS},y \gtrsim 10^{-4}$,
which implies that $\Lambda_{\mathrm{H}}<200$ TeV.\footnote{A large $\Lambda_{\mathrm{H}}$,  which is realized by the small couplings $y$ and $\lambda_{HS}$, does not necessarily mean a heavy $S$ as shown in Fig.~\ref{MassSpectrum}. Therefore, even if $\Lambda_{\mathrm{H}}$ is large, the correction to the Higgs mass coming from the internal $S$ loop can be small.}

\begin{figure}[t]
\begin{center}
\includegraphics[width=4in]{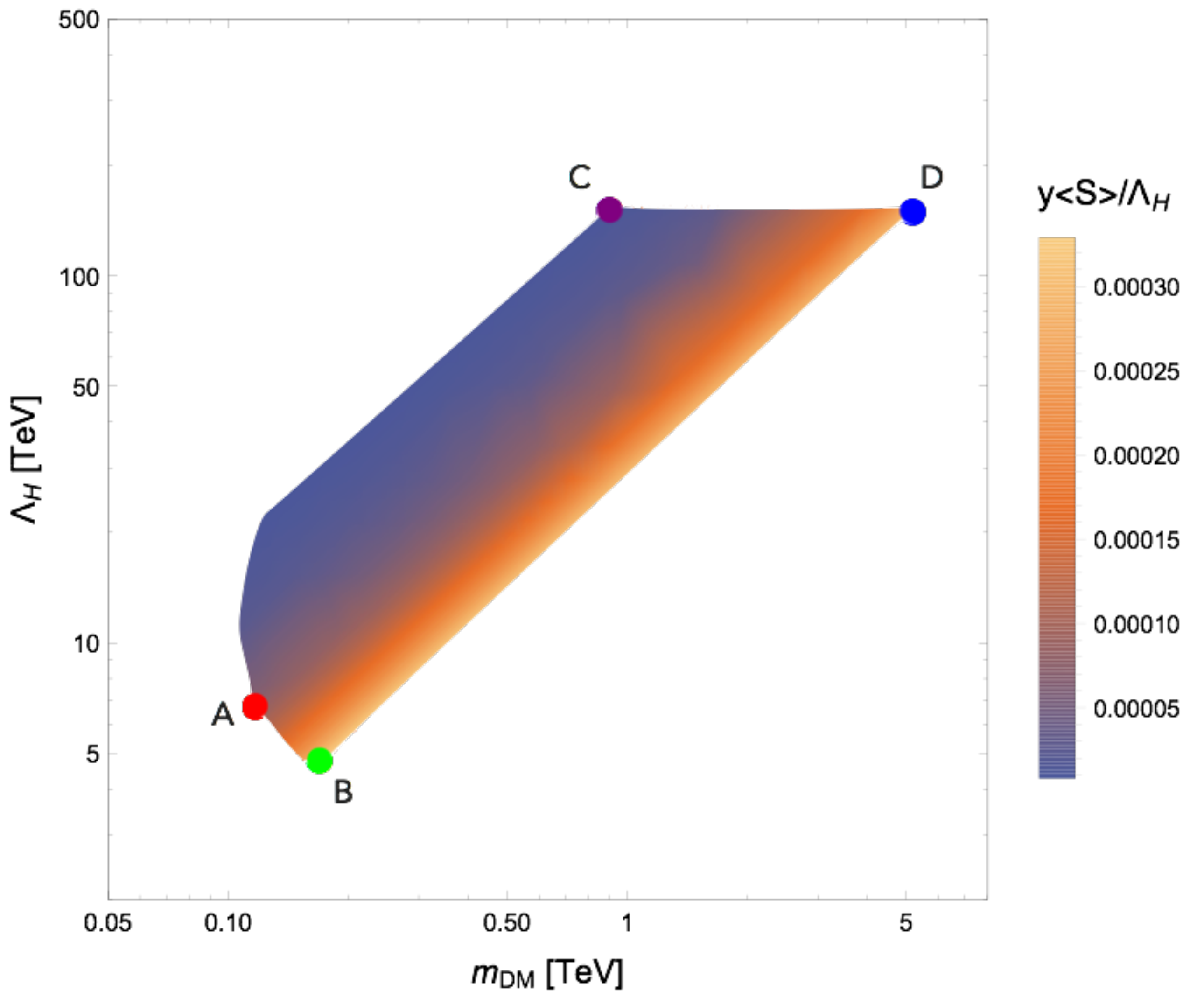}
\vspace{-2mm}
\caption{The hidden QCD scale $\Lambda_{\mathrm{H}}$ against the DM mass $m_{\mathrm{DM}}$. In the colored region,
$\left<h\right>=246~\mathrm{GeV}$, $m_h=125.09\pm 0.24~\mathrm{GeV}$,  $\xi_{1}^{(1)}>0.99$ ($h$-$S$ mixing), $m_S\simeq 2m_{\mathrm{DM}}$, and the perturbativity and stability constraint (\ref{scalars}) are satisfied.
We assumed the case for $\lambda_{HS},y \gtrsim 10^{-4}$.
The color strength indicates the value of $y\left<S\right>/\Lambda_{\mathrm{H}}$ which is a measure of how the chiral symmetry is explicitly broken. 
The colored points are the benchmark points; Cases A (red), B (green), C (purple), and D (blue) are defined in Table \ref{CaseABCD}.}
\label{ABCDmc}
\end{center}
\end{figure}

In Fig.~\ref{ABCDmc} we show the area in the $m_{\mathrm{DM}}$-$\Lambda_{\mathrm{H}}$ plane, in which we obtain the VEV of the Higgs field $\left<h\right>=246~\mathrm{GeV}$, the correct Higgs mass $m_h=125.09\pm 0.24~\mathrm{GeV}$ \cite{Olive:2016xmw}, $h$-$S$ mixing $\xi_{1}^{(1)}>0.99$ \cite{Olive:2016xmw}, and 
the resonance condition $m_S\simeq2m_{\mathrm{DM}}$ (to realize the correct DM relic abundance).
Note that the mass of the mediator $S$ is bounded, 
because $\lambda_{S}$ is bounded as discussed above.
Consequently, because of the resonance condition $m_S\simeq2m_{\mathrm{DM}}$,
the DM mass is bounded, too.
Similarly, $\Lambda_{\mathrm{H}}$ is bounded,
because $\lambda_{HS}$ is bounded from above (\ref{scalars})
and from below due to our parameter choice $\lambda_{HS}>10^{-4}$.
The colored points A, B, C and D in Fig.~\ref{ABCDmc} are our benchmark points.

The chosen four benchmark points are named Case A, B, C, and D:
the set of the input parameter values $(\lambda_{H},\lambda_{HS},\lambda_{S},y)$, 
along with the output values of $m_{\rm DM}$, $\Lambda_{\mathrm{H}}$, and
 $y\left<S\right>/\Lambda_{\mathrm{H}}$ for each benchmark case is given in Table \ref{CaseABCD}.
Under $y\lesssim 0.006$ and (\ref{scalars}), Cases A and B are located 
as close to the EW scale as possible and for C and D in an opposite way.\footnote{There exists the Higgs threshold between Cases A and B, 
which means the decay channel of the mediator $S$ to two Higgs particles is forbidden only for the Case A.
This might become a benchmark point for a future collider search.}
We regard the normalized current quark mass $y\left<S\right>/\Lambda_{\mathrm{H}}$ as the characterization for the explicit chiral symmetry
breaking.
Their values should be compared with that of QCD, i.e. $m_u/\Lambda_{\mathrm{QCD}}\sim 6\times 10^{-3}$ (in the NJL model).

%=============   \ref{ABCDmc}   ==============%
%=================================%

%%%%%%%%Table %%%%%%%%
 \begin{table}
  \caption{Four benchmark points, Cases A--D, which are defined by the values of $(\lambda_H,\lambda_{HS},\lambda_S,y)$, where $m_{\mathrm{DM}},~\Lambda_{\mathrm{H}}$, and $y\left<S\right>/\Lambda_{\mathrm{H}}$ 
  are displayed for each case.\\  }
  \label{CaseABCD}
  \centering
  \begin{tabular}{|c||c|c|c|c|}
    \hline
    ~~Case~~ & ~~$(\lambda_H,\lambda_{HS},\lambda_S,y)$~~  & ~~$m_{\mathrm{DM}}$ [TeV]~~  & ~~$\Lambda_{\mathrm{H}}$ [TeV]~~& ~~$y\left<S\right>/\Lambda_{\mathrm{H}}$~~  \\
  \hline
   A &  ~~$(~0.140,~0.050,~0.054,~8.57\times10^{-4})~~$ &  $0.117$ &   $6.84$  &   ~~$7.30\times 10^{-6}$~~\\
       \hline
   B &  ~~$(~0.138,~0.098,~0.230,~3.60\times10^{-3})~~$ &  $0.170$ &   $4.87$  &   ~~$3.05\times 10^{-5}$~~\\
       \hline
   C &  ~~$(~0.129,~0.0001,~0.007,~1.07\times10^{-4})~~$ &  $0.906$ &   $153.1$  &   ~~$8.73\times 10^{-6}$~~\\
       \hline
   D &  ~~$(~0.130,~0.0001,~0.230,~3.55\times10^{-3})~~$ &  $5.20$ &   $152.5$  &   ~~$2.90\times 10^{-5}$~~\\
       \hline
  \end{tabular}
\end{table}
%%%%%%%%Table %%%%%%%%

\begin{figure}[t]
\begin{minipage}{0.49\hsize}
\begin{center}
\includegraphics[width=3.2in]{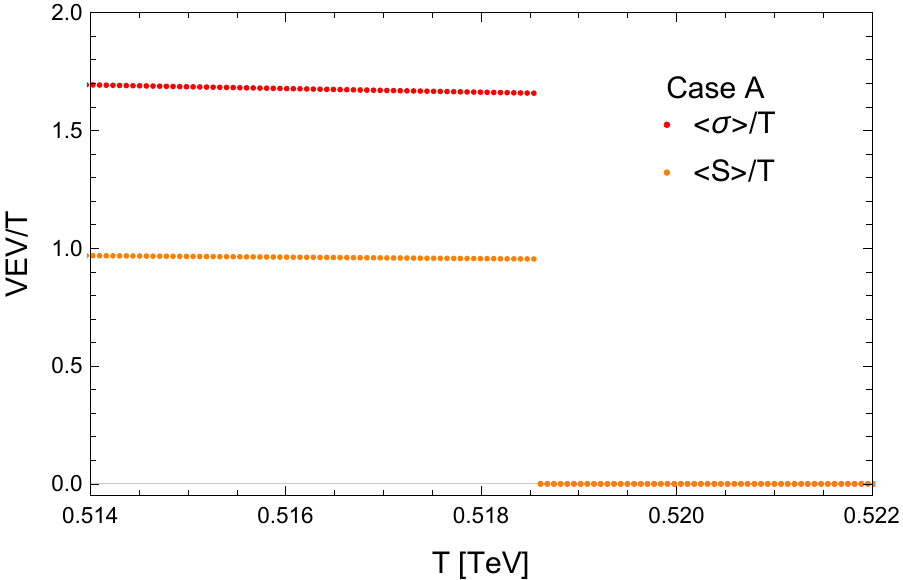}
\end{center}
\end{minipage}
\begin{minipage}{0.49\hsize}
\begin{center}
\includegraphics[width=3.1in]{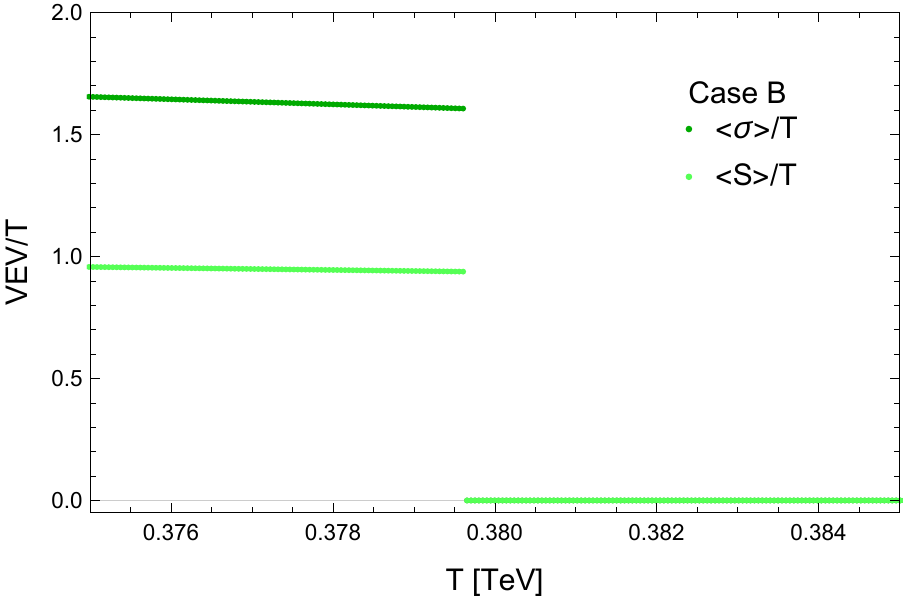}
\end{center}
\end{minipage}
\begin{minipage}{0.49\hsize}
\begin{center}
\vspace{+2mm}
\includegraphics[width=3.2in]{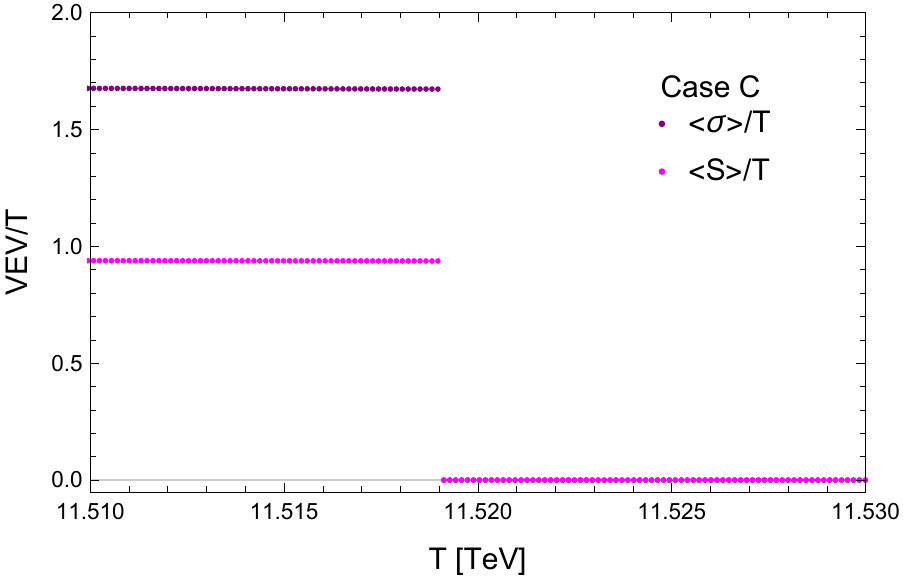}
\end{center}
\end{minipage}
\begin{minipage}{0.49\hsize}
\begin{center}
\vspace{+2mm}
\includegraphics[width=3.2in]{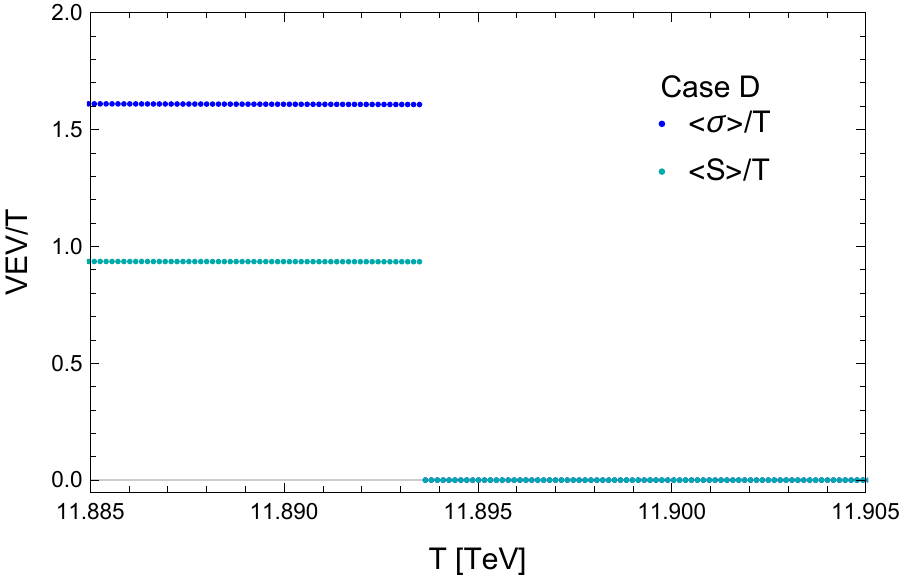}
\end{center}
\end{minipage}
\caption{The temperature dependence of $\left<\sigma\right>/T$ (dark colored) and $\left<S\right>/T$ (light colored) for each benchmark point. Cases A (top left), B (top right), C (bottom left), and D (bottom right) are defined in Table \ref{CaseABCD}. }
\label{PTABCD}
\end{figure}

In Fig.~\ref{PTABCD} we show the temperature dependence of $\left<\sigma\right>/T$ and $\left<S\right>/T$ near the critical temperature for each benchmark point.
It can be seen that
a first-order phase transition appears in all the cases.
We also see that  $\sigma$ and $S$ undergo the phase transition at the same time.
Moreover, 
the phase transition in Case C appears for a slightly lower temperature compared with D even though $\Lambda^{\mathrm{C}}_{\mathrm{H}}>\Lambda^{\mathrm{D}}_{\mathrm{H}}$.
The reason is that the explicit chiral symmetry breaking, whose strength is expressed by $y\left<S\right>/\Lambda_{\mathrm{H}}$,
influences not only the mass of DM but also the critical temperature.
Since the chiral phase transition in the hidden sector
occurs in the two-dimensional space $(S,\sigma)$,
we need to deal with quantum tunneling in the two-dimensional space to calculate the GW spectrum.

%=============   \ref{PTABCD}   ==============%
%=================================%

%=======================================================================%
%=======================================================================%
\section{Bubbles from Hidden QCD Tunneling}\label{Bubbles from the Tunneling
in the Hidden Sector}
Cosmological tunneling has been studied in \cite{Coleman:1977py,Callan:1977pt,Linde:1981zj}.
The probability of the bubble nucleation per unit volume per unit time is given by
 \begin{align}
 \label{Gamma}
\Gamma=A(t) \exp \left[ -S_E(t) \right],
 \end{align}
where $S_E$ is the Euclidean action. 
At a high temperature,
the Euclidean action can be replaced by
$S_E=S_3/T$ because of the periodicity of $S_E$ in the Euclidean time,
where $S_3$ is the corresponding
three-dimensional Euclidean action \cite{Linde:1981zj}.
The bubbles can percolate when the probability of  the bubble nucleation per unit volume and 
time is of order one.
Since the prefactor $A$ in (\ref{Gamma}) is $A(T)\propto T^4$ \cite{Linde:1981zj},
we can translate this condition as
\begin{align}
\label{PTcondition}
 \left. \frac{\Gamma}{H^4}\right|_{t=t_{\text{t}}} \simeq 1   ~~\rightleftarrows~~ \frac{S_3 (T_{\text{t}})}{T_{\text{t}}} = 4\ln \left( \frac{T_{\text{t}}}{H_{\text{t}}}\right),
\end{align} 
where $H_{\mathrm{t}}$ is the Hubble parameter at the transition temperature $T_{\mathrm{t}}$.

The bubble dynamics can be characterized by two parameters,
namely, $\alpha$ and $\beta$ at $T_{\text{t}}$ \cite{Grojean:2006bp}:
$\alpha$ expresses how much energy the phase transition releases,
 while $\beta^{-1}$ expresses how long its phase transition takes.
These parameters are essential for computing the GW signal from the cosmological phase transition \cite{Grojean:2006bp}.
The parameter $\alpha$ is defined as 
\begin{align}
\label{alpha}
  \alpha  & \equiv  \left. \frac{\epsilon }{\rho_{\text{rad}}}\right|_{T=T_{\text{t}}},
\end{align}
which is the ratio of the latent heat $\epsilon$  liberated at the phase transition 
to the thermal energy density $\rho_{\text{rad}}(T_{\mathrm{t}})=(\pi^2/30)g_*(T_{\mathrm{t}})T_{\mathrm{t}} ^4$ in the symmetric phase. 
The latent heat can be computed from the effective potential at finite temperature as
\begin{align}
\label{}
  \epsilon (T) & \equiv -\Delta V_{\text{EFF}}(T)+T\frac{\partial \Delta V_{\text{EFF}}(T)}{\partial T},
\end{align}
where $\Delta V_{\text{EFF}}(T)$ is the difference of the effective potential between the true and false vacuum.
The parameter $\beta$ is defined as
\begin{align}
\label{}
  \beta  & \equiv  -\left. \frac{d S_E}{dt}\right|_{t=t_{\text{t}}}\simeq  \left. \frac{1}{\Gamma}\frac{d\Gamma}{dt}\right|_{t=t_{\text{t}}}.
\end{align}
Using $H_{\mathrm{t}}$, we can redefine 
a dimensionless parameter $\tilde{\beta}$ as
 \begin{align}
\label{tildebeta}
  \tilde{\beta}  & \equiv  \frac{\beta}{H_{\text{t}}}=T_{\text{t}}\left. \frac{d}{dT}\left(\frac{S_3(T)}{T}\right)\right|_{T=T_{\text{t}}}.
\end{align}
In the following subsections we apply above the general formula
(\ref{PTcondition})--(\ref{tildebeta})
to compute the parameters $(T_{\mathrm{t}},\alpha, \tilde{\beta})$
for our concrete problem,
and we estimate the corresponding GW signal.

%=======================================================================%
\subsection{Bubble Nucleation and Tunneling Parameters}
In order to discuss the bubble nucleation which stems from the first-order chiral phase transition,
we need to calculate $S_3$.
For this purpose we use the effective Lagrangian for the mean field $\sigma$. 
However, the mean field $\sigma$ cannot describe tunneling at a tree level,
because its kinetic term is absent at the tree level.
Hence we compute its kinetic term from the two-point function $\Gamma_{\sigma\sigma}$ at the one-loop level, which is given in (\ref{CPEVENscalar}). 
First we discuss the zero-temperature case and 
define the field renormalization constant $Z_{\sigma}$ for the $\sigma$ field as 
$$
\Gamma_{\sigma\sigma}(p^2) =\Gamma_{\sigma\sigma}(0)+ Z^{-1} _{\sigma}(S,\sigma) p^2 + \mathcal{O}(p^4),
$$
where
\begin{align}
\nonumber
   Z^{-1} _{\sigma}(S,\sigma)  & = -\left( 1-\frac{G_D}{4G^2}\sigma \right)^2 3 n_c \left. \frac{d}{dp^2}I_{\varphi^2}(p^2,M;\Lambda_{\mathrm{H}})\right|_{p^2=0}.
\end{align}
Thus the effective Lagrangian for the $\sigma$ field at zero temperature is
 \begin{align}
 \label{Lagrangian for sigma}
\mathcal{L}_{\sigma} &=\frac{ Z^{-1}_{\sigma}(S,\sigma)}{2}\partial_{\mu}\sigma\partial^{\mu}\sigma-V_{\mathrm{eff}}(S, \sigma),
 \end{align}
 where $V_{\mathrm{eff}}(S,\sigma)=V^{h\rightarrow0}_{\text{SM}+S}(S)+V_{\text{NJL}} (S,\sigma)$
 [$V_{\text{NJL}} (S,\sigma)$ is given in (\ref{Potential0T})].
Note that the field renormalization constant $Z^{-1}_{\sigma}$ at the symmetric phase ($S=\sigma=0$) diverges (see the Appendix).
This is expected, because the composite state $\sigma$ disappears in the symmetric phase.

\begin{figure}[t]
\begin{center}
\includegraphics[width=3.8in]{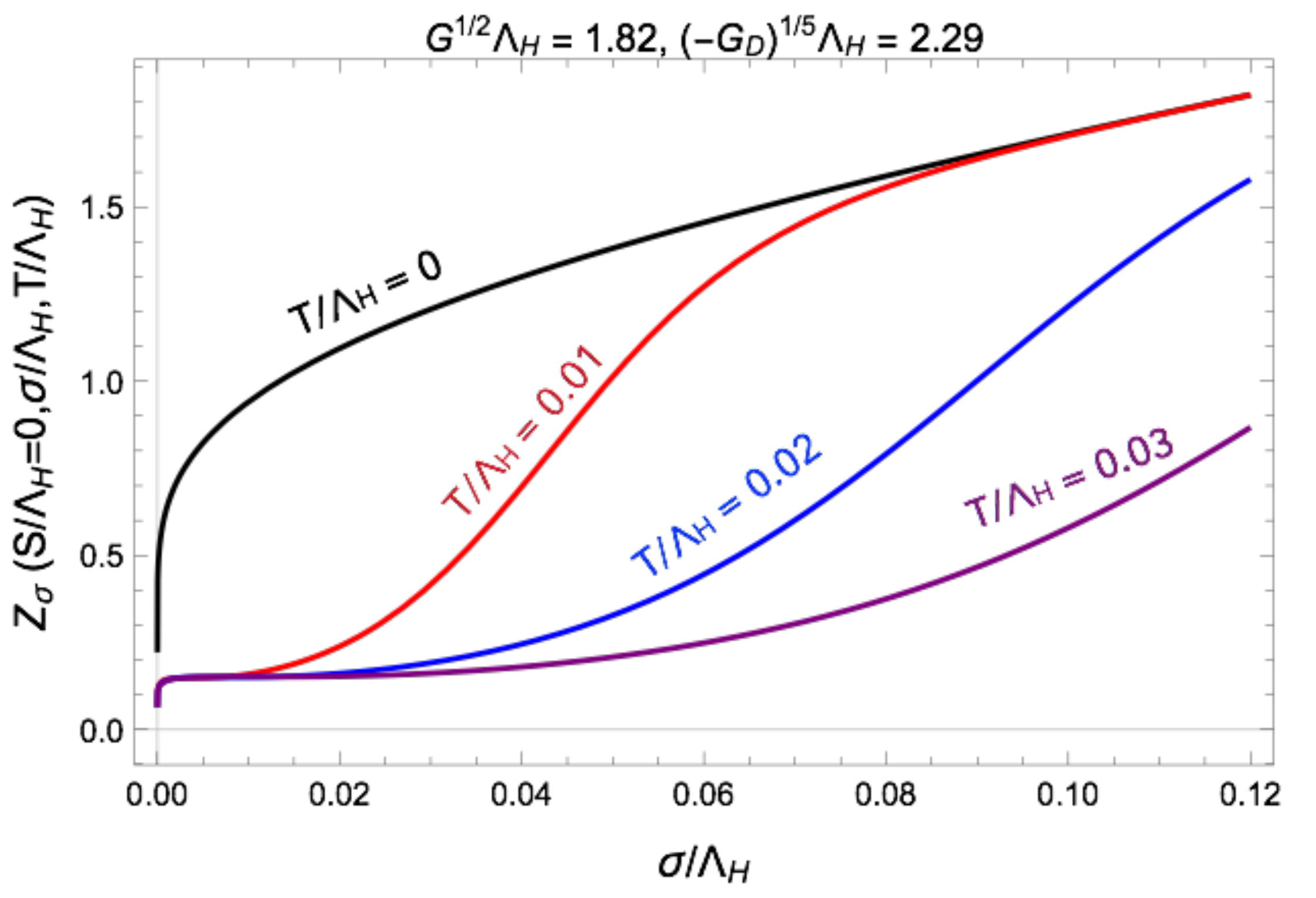}
\end{center}
\caption{The $\sigma$ field dependence of the field renormalization constant $Z_{\sigma}(S=0,\sigma,T)$ for $T/\Lambda_{\mathrm{H}}=0~\mathrm{(black)},~0.01~\mathrm{(red)},~0.02~\mathrm{(blue)},~\mathrm{and}~0.03~\mathrm{(purple)}$.}
\label{Zsig}
\end{figure}

As mentioned in the previous section, hidden QCD tunneling should occur in the two-dimensional field space and
 could be described by the three-dimensional Euclidean action 
 \begin{align}
  \label{actionS}
S_3(T)=\int d^3x \left[\frac{Z^{-1} _{\sigma}(S,\sigma,T)}{2}
\left( \partial_i \sigma\right)^2+\frac{1}{2}
\left( \partial_i S\right)^2+V_{\mathrm{EFF}}(S,\sigma,T) \right].
\end{align}
The field renormalization constant at finite temperature is 
computed in the Appendix and found to be 
 \begin{align}
\nonumber
 & ~~~~~Z^{-1}_{\sigma} (S,\sigma,T ) \\
\label{}
& = \frac{3n_c}{8\pi^2}\left(1- \frac{G_D}{4G^2}\sigma\right)^2\left[\ln \left(1+\frac{\Lambda^2_{\mathrm{H}}}{M^2}\right)+\frac{\Lambda^2_{\mathrm{H}}M^2}{(\Lambda^2_{\mathrm{H}}+M^2)^2}-32\pi^2\left( A_F(u^2)-B_F(u^2)\right)\right] ,
\end{align}
where $u= M/T$, and $M$, $A_F(u^2)$, and $B_F(u^2)$ are given in Eqs.~(\ref{M}), (\ref{AFT}), and (\ref{BFT}), respectively.
In Fig.~\ref{Zsig} we show the field dependency of the field renormalization constant $Z_{\sigma}(S,\sigma,T )$ for $S=0$ and $T/\Lambda_{\mathrm{H}}=0,~0.01,~0.02,~\mathrm{and}~0.03$, which corresponds to the black, red, blue, and purple line, respectively.
As shown in Fig.~\ref{Zsig},
the field renormalization constant $Z_{\sigma}(S,\sigma,T)$ vanishes in  the symmetric phase.
The  $\mathrm{O}(3)$ symmetric bounce solution can be obtained by solving the equations of motion
 \begin{align}
      \label{dfeqs}
  \frac{d^2 \sigma}{dr^2} + \frac{2}{r}\frac{d \sigma}{dr}
  +\frac{1}{2}\frac{\partial \ln Z_{\sigma}(S,\sigma,T)}{\partial \sigma}\left(\frac{d \sigma}{dr}\right)^2 & =Z_{\sigma}(S,\sigma,T)\frac{\partial V_{\mathrm{EFF}}(S,\sigma,T) }{\partial \sigma},\\
   \label{dfeqS}
  \frac{d^2 S}{dr^2} + \frac{2}{r}\frac{d S}{dr} 
-\frac{1}{2}\frac{\partial  Z^{-1}_{\sigma}(S,\sigma,T)}{\partial S}\left(\frac{d \sigma}{dr}\right)^2& =\frac{\partial V_{\mathrm{EFF}}(S,\sigma,T) }{\partial S},
\end{align}
where $r=\left(x^2_1+x^2_2+x^2_3\right)^{1/2}$.
The boundary conditions are 
\begin{align}
\label{boundary}
\left.\frac{d\sigma}{dr}\right|_{r=0}=0,~~\left.\frac{dS}{dr}\right|_{r=0}=0,~~\lim_{r\rightarrow \infty}\sigma(r)=0,~~\lim_{r\rightarrow \infty}S(r)=0,
\end{align}
where the coordinate of the symmetric minimum (false vacuum) of the potential
is chosen at the origin of the $\sigma$-$S$ space.
Note that the field renormalization constant $Z_{\sigma}(S,\sigma,T)$ does not depend explicitly on $r$ but also depends on the fields.

%=============   \ref{Zsig}   ==============%
%=================================%

%=======================================================================%
\subsection{Computation of Multi-dimensional Bounce Solution }
In the one-dimensional case we can obtain a bounce solution by using 
the so-called overshooting/undershooting method \cite{Apreda:2001us}. 
However,  this is a cumbersome method in the multidimensional case,
because two initial conditions have to be simultaneously  fine-tuned. 
Instead, we here employ  an approach similar to the path deformation method \cite{Wainwright:2011kj}.

The bounce solution is unique. That is, $\sigma(r)$ and $S(r)$,
which satisfy the differential equations (\ref{dfeqs})  and (\ref{dfeqS}) 
with the boundary conditions (\ref{boundary}), are a unique function 
of $r$. If we assume that $\sigma(r)$ is an invertible function
for $r  \in [0,\infty)$, then there exists a unique inverse of $\sigma$,
which we denote by $\sigma^{-1}$. That is,  $\sigma^{-1} \circ \sigma$
is the identity function, or $r= \sigma^{-1} (\sigma(r))$.
Because of this assumption,  $S$
can be regarded as  a function of $\sigma$, i.e., $S(\sigma)$.\footnote{We use the same symbol $S$ for the functions of $r$ and $\sigma$.}
Therefore, (\ref{dfeqs}) and (\ref{dfeqS}) can be written as, respectively,
\begin{align}
      \label{dfeqs2}
  \frac{d^2 \sigma}{dr^2} + \frac{2}{r}\frac{d \sigma}{dr} 
 + \frac{1}{2}\frac{\partial \ln Z_{\sigma}(S(\sigma),\sigma,T)}{\partial \sigma}\left(\frac{d \sigma}{dr}\right)^2& =
  F_{\sigma}(S(\sigma),\sigma),\\
        \label{dfeqS2}
  \frac{d^2S}{d\sigma^2}\left(\frac{d\sigma}{dr}\right)^2+\left(\frac{dS}{d\sigma }\right)\left(\frac{d^2 \sigma}{dr^2} + \frac{2}{r}\frac{d \sigma}{dr} \right)
  -\frac{1}{2}\frac{\partial  Z^{-1}_{\sigma}(S(\sigma),\sigma,T)}{\partial S}\left(\frac{d \sigma}{dr}\right)^2  &=F_{S}(S(\sigma),\sigma),
\end{align}
where $F_{\sigma}(S(\sigma),\sigma)$ and 
$F_{S}(S(\sigma),\sigma)$ are the rhs of (\ref{dfeqs})
and (\ref{dfeqS}), respectively, and we have suppressed the $T$ dependence
of $F_{\sigma}(S(\sigma),\sigma)$ and $F_{S}(S(\sigma),\sigma)$.
The point is that if $S(\sigma)$ is given, then 
(\ref{dfeqs2}) is a one-dimensional differential equation and hence
can be solved by applying the overshooting/undershooting method.
If $S(\sigma)$ is the  true solution of the problem, it should satisfy (\ref{dfeqS2})
with $\sigma(r)$ obtained from (\ref{dfeqs2}) as well, which means
that
\be
\label{NS}
N(r) &=&0
\ee
is also satisfied, where
\begin{align}
N(r) &=\frac{d^2S}{d\sigma^2}(r)
\left(\frac{d\sigma}{dr}(r)\right)^2+\frac{dS}{d\sigma }(r)
 F_{\sigma}(S,\sigma)(r)-F_{S}(S,\sigma)(r)\nn\\
 \label{NS-def}
 & -\frac{1}{2}\left(\frac{d\sigma}{dr}(r)\right)^2
 \left(
 \frac{\partial Z^{-1}_{\sigma}(S,\sigma,T)}{\partial S}(r)
 +\frac{dS}{d\sigma }(r) \frac{\partial \ln Z_{\sigma}(S,\sigma,T)}{\partial \sigma}(r)
 \right).
\end{align}
Since the one-dimensional differential equation (\ref{dfeqs2})
 for a given path $S(\sigma)$
can be simply solved, our task is  to find $S(\sigma)$ which satisfies (\ref{NS}).  We do this in an iterative way.
We start with a linear function $S_0(\sigma)$, which connects 
the true and false vacuum:
\begin{align}
\label{S0}
S_{0}(\sigma)=\frac{S^B-S^S}{\sigma^B-\sigma^S}\left(\sigma-\sigma^S\right)+S^S,
\end{align} 
where $(S^{B,S},\sigma^{B,S})$ 
(with $S^{S}=\sigma^{S}=0$ ) are the positions of  the true and false vacuum, respectively.
Then we solve  (\ref{dfeqs2}) with the path $S(\sigma)=S_0(\sigma)$ and
denote the bounce solution by $\sigma_0(r)$.
Note that $\sigma_0(0)$ is no longer $\sigma^B$, so that the end point
of $S_0(\sigma)$
on the true vacuum side is no longer $S^B$, i.e.
$S_0 (\sigma_0(0)) \neq S^B$.
Next  we  compute  the rhs of (\ref{NS-def})  using 
$\sigma_0(r)$ and $S_0(\sigma_0(r))$ for $\sigma$ and $S(\sigma)$, respectively,
and we denote it by $N_0(r)$.
Since $S_0(\sigma_0(r))$ is not the true solution of the problem,
$N_0(r)$  does not vanish. 
Knowing $N_0(r)$, we have to define the next step
of the iteration:
\be
S_1(\sigma) &= & S_0(\sigma)+\Delta S_0(\sigma).
\ee
To proceed we assume that not only the true solution
$\sigma(r)$ but also $\sigma_0(r)$ is an invertible function, so that
$N_0(r)$ can be written as a function of $\sigma$, i.e.,
\be
\hat{N}_0(\sigma) & \equiv & N_0(r=\sigma_0^{-1}(\sigma)).
\label{Nhat}
\ee
Note that because of the $\sigma$ and $S$ dependence of 
$Z_\sigma(S,\sigma,T)$ (partly shown in  Fig.~\ref{Zsig}) and also of
$V_{\rm EFF}(S,\sigma,T)$, $\hat{N}_0(\sigma)$
vanishes at the false vacuum,
 i.e., at $\sigma=0$ ($S$ also vanishes at $\sigma=0$).
 Further, if $\hat{N}_0(\sigma)$ vanishes at some nonzero values of
 $\sigma$, the deformation $\Delta S_0(\sigma)$ should also vanish
 at these values of  $\sigma$.
 This brings us to assume that  $\Delta S_0(\sigma)$ is proportional
to $\hat{N}_0(\sigma)$.
Therefore, the path $S_{i+1}(\sigma)$ 
in the $(i+1)$ th step  can be defined as
\be
S_{i+1}(\sigma) &= & S_i(\sigma)+k\hat{N}_i(\sigma),
\ee
where $k$ is the step size, and
$\hat{N}_i(\sigma) = N_i (r=\sigma_{i}^{-1}(\sigma))$. Note that
$S_{i+1}(\sigma)$ satisfies
the boundary condition $\lim_{\sigma \to 0}S_{i+1}(\sigma)=0$.
To obtain $\sigma_{i+1}(r)$, the initial value of $\sigma_{i+1}(0)$
has to be fine-tuned in such away that 
$\left.d \sigma_{i+1}(r)/d r\right|_{r=0}=0$ and
$\lim_{r\to \infty} \sigma_{i+1}(r)=0$ are satisfied.
 If $\left.d \sigma_{i+1}(r)/d r\right|_{r=0}=0$ is satisfied,
$\left.d S_{i+1}(\sigma_{i+1}(r))/d r\right|_{r=0}=0$ is automatically satisfied.
Since  $\sigma_{i+1}(0)$ is different  from
$\sigma_{i}(0)$, the end point
of $S_{i+1}(\sigma)$
on the true vacuum side is also moved
to $S_{i+1}(\sigma_{i+1}(0))$.

\begin{figure}[t]
\begin{minipage}{1\hsize}
\begin{center}
\includegraphics[width=4.2in]{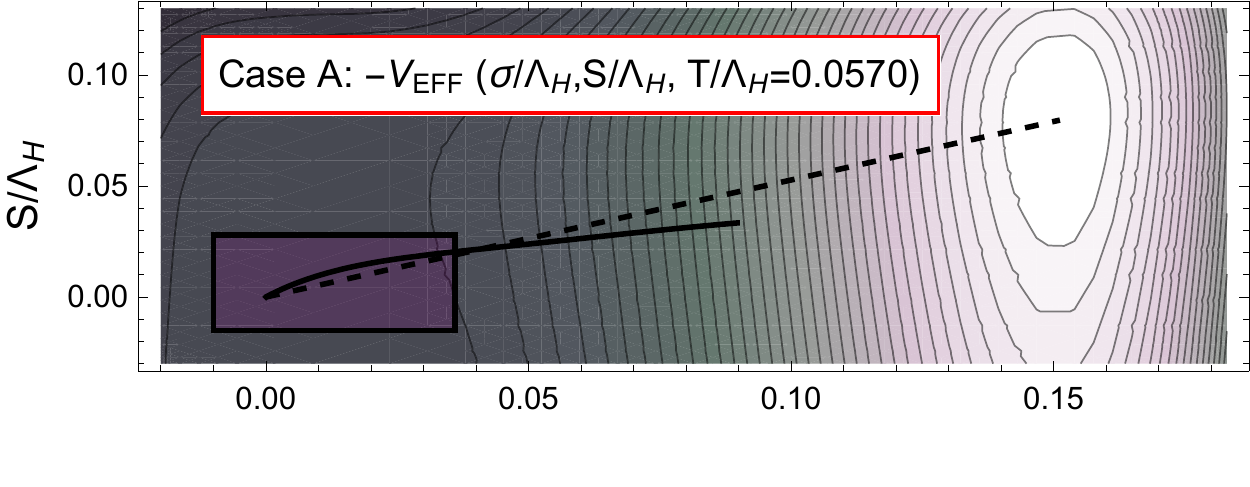}
\end{center}
\end{minipage}
\begin{minipage}{1\hsize}
\begin{center}
\includegraphics[width=4.2in]{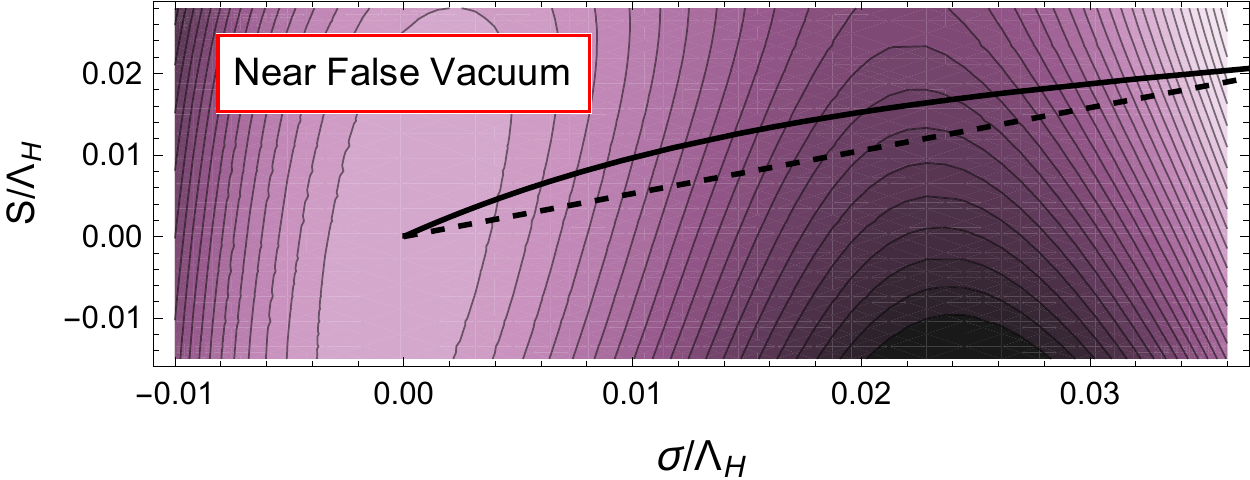}
\end{center}
\end{minipage}
\caption{Top: The contour plot of the effective potential $V_{\mathrm{EFF}}$ in (\ref{VEFF}) with  $T/\Lambda_{\mathrm{H}}=0.0570$ for Case A,
 defined in Table \ref{CaseABCD}. The black dashed line stands for 
  the initial path $S_0(\sigma)$ and the black solid line is 
  the  path $S_{15}(\sigma)$ with  $|k\hat{N}_{15}(\sigma)|/
S_{15}(\sigma) < 10^{-2}$.  Bottom: 
The region enclosed by the box near the false vacuum in the top figure
is zoomed.}
\label{Path}
\end{figure}

\begin{figure}[t]
\begin{center}
\includegraphics[width=3.6in]{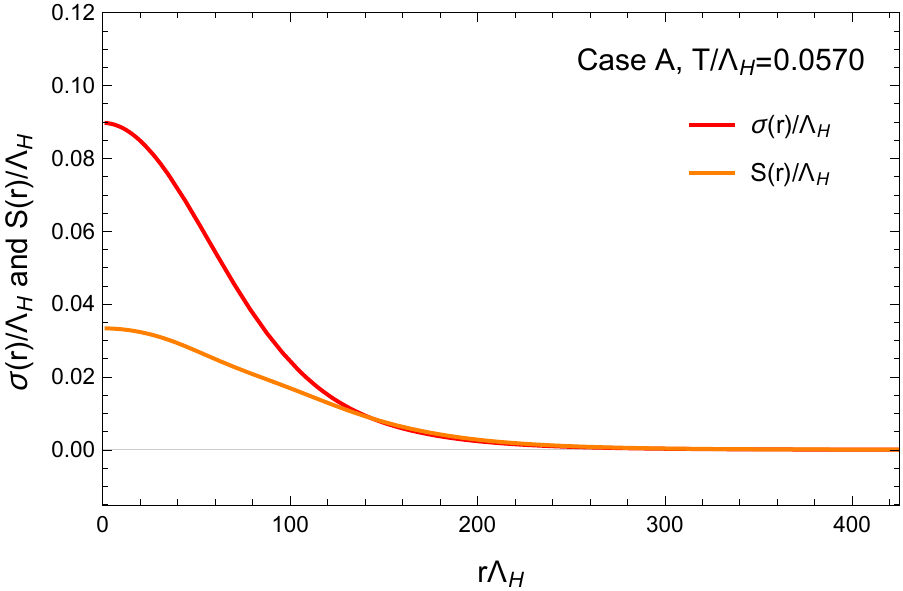}
\end{center}
\caption{The bounce solution for Case A
with $T/\Lambda_{\mathrm{H}}=0.0570$.
The red line stands for $\sigma(r)/\Lambda_{\mathrm{H}}$, 
and  the  orange one for $S(r)/\Lambda_{\mathrm{H}}$,
 which correspond to the path $S_{15}(\sigma)$ (black solid line) shown in Fig.~\ref{Path}. }
\label{MultiBounce}
\end{figure}

Since the assumptions we made above cannot be rigorously
justified, there is no guarantee that the steps converge to the 
true solution of the problem. In fact, if we choose the wrong sign for 
$k$, steps diverge or do not converge.
We have checked our method for a number of examples and found
that once we use an appropriate  sign and size for $k$,
the steps can converge, where we approximate 
the path $S_{i}(\sigma)$ (which is obtained numerically)
 with a fifth-degree polynomial in $\sigma$ as in \cite{Beniwal:2017eik}.
In Fig.~\ref{Path}, we present the numerical solution 
$S_{15}(\sigma)$(black solid line) with $|k\hat{N}_{15}(\sigma)|/
S_{15}(\sigma) < 10^{-2}$
obtained  from  $S_0(\sigma)$ (black dashed line)
 in the two-dimensional field space at  $T=0.390$ TeV
 [below the critical temperature $T=0.519$ TeV as shown in Fig.~\ref{PTABCD} (top left)] for Case A.\footnote{$|k\hat{N}_{i}(\sigma)|/
S_{i}(\sigma) < 10^{-2}$ is not satisfied for $i<15$.}
The corresponding bounce solution
 as a function of $r$ is shown in Fig.~\ref{MultiBounce}.
The Euclidean action (\ref{actionS}) obtained from the bounce solution
is $S_3(T)/T=148.2$,
where the difference of $S_3(T)/T$ between the 14th and 15th steps is less than a few percent.
Computing $S_3(T)/T$ for each temperature as in the above method,
we can find the transition temperature $T_{\mathrm{t}}$ from the condition (\ref{PTcondition}),
which is used for the determination of tunneling parameters $\alpha$ and $\tilde{\beta}$ given in Eqs.~(\ref{alpha}) and (\ref{tildebeta}).

%=============   \ref{Path}   ==============%
%=================================%

%=============   \ref{MultiBounce}   ==============%
%=================================%

%=======================================================================%
\subsection{Tunneling Parameters for the Benchmark Points}

The GW spectrum produced by a  first-order phase transition can be 
characterized by the released  energy and its duration time,
and it is known that they can be parametrized by the set of the parameters 
$(T_{\mathrm{t}},\alpha, \tilde{\beta})$. 
The results for the benchmark points are given in Table \ref{GWparas}.
We see from Table \ref{GWparas} that
 $\alpha$ and $ \tilde{\beta}^{-1}$ for Cases A and C are larger than
 those for Cases B and D. Recalling  the parameter values for the benchmark points
(Table \ref{CaseABCD}), we can infer that the smaller the explicit chiral symmetry
breaking (the smaller $y$) is, the larger  $\alpha$ and $ \tilde{\beta}^{-1}$ are.
This suggests  that the parameters of the model can be constrained
if the GW is measured with a certain accuracy.

%%%%%%%%Table %%%%%%%%
 \begin{table}
  \caption{The parameters $(T_{\mathrm{t}},\alpha,\tilde{\beta})$ for benchmark points defined in Table \ref{CaseABCD}. The transition temperature $T_{\mathrm{t}}$, the ratio of the latent heat to the
   thermal energy density $\alpha$, and the dimensionless inverse duration time $\tilde{\beta}$ are defined by Eqs.~(\ref{PTcondition}), (\ref{alpha}), and (\ref{tildebeta}), respectively.\\}
  \label{GWparas}
  \centering
  \begin{tabular}{|c||c|c|c|}
    \hline
    ~~Case~~ &  ~~$T_{\mathrm{t}}$ [TeV]~~  & ~~$\alpha$ ~~& ~~$\tilde{\beta}$~~  \\
  \hline
   A &  ~~$0.387$~~ & ~~ $0.288$ ~~ &   ~~$8.24\times 10^2$~~\\
       \hline
   B &  ~~$0.306$~~ & ~~ $0.223$ ~~ &   ~~$14.86\times 10^2$~~\\
       \hline
   C &  ~~$8.731$~~ & ~~ $0.310$ ~~ &   ~~$7.15\times 10^2$~~\\
       \hline
   D &  ~~$9.480$~~ & ~~ $0.232$ ~~ &   ~~$13.29\times 10^2$~~\\
       \hline
  \end{tabular}
\end{table}
%%%%%%%%Table %%%%%%%%

%=======================================================================%
%=======================================================================%
\section{Signal From the Hidden Sector QCD}\label{Signal From Hidden QCD}
Finally we come to our  main purpose: to check the testability of the GW background produced by the first-order phase transitions in the hidden sector.
There coexist three processes contributing to  the stochastic GW background
spectrum:
 \begin{align}
h^2\Omega_{\mathrm{GW}}=h^2\Omega_{\mathrm{\varphi}}+h^2\Omega_{\mathrm{sw}}+h^2\Omega_{\mathrm{turb}},
 \label{omeGW}
\end{align}
where $h$ is the dimensionless Hubble parameter, $\Omega_{\mathrm{\varphi}}$ stands for the scalar field contribution from collisions of bubble walls \cite{Kosowsky:1991ua,Kosowsky:1992rz,Kosowsky:1992vn,Kamionkowski:1993fg,Caprini:2007xq,Huber:2008hg}, 
$\Omega_{\mathrm{sw}}$ for the contribution from
sound waves in plasma after the bubble collisions \cite{Hindmarsh:2013xza,Giblin:2013kea,Giblin:2014qia,Hindmarsh:2015qta},
 and $\Omega_{\mathrm{turb}}$ for the contribution 
 from magnetohydrodynamic (MHD) turbulence in plasma \cite{Caprini:2006jb,Kahniashvili:2008pf,Kahniashvili:2008pe,Kahniashvili:2009mf}.
Following \cite{Caprini:2015zlo},
each contribution is given for a given set of the parameters $( T_{\mathrm{t}},\alpha,\tilde{\beta})$ with
the velocity of bubble wall $v_w$ and the $\kappa_{\varphi}, \kappa_{v}$, and $\kappa_{\mathrm{turb}}$ which are the fraction of vacuum energy, respectively, converted into gradient energy of scalar field, bulk motion of the fluid, and MHD turbulence.

\begin{itemize}
%===============
  \item Scalar field contribution $\Omega_{\varphi}$:
 \begin{align}
h^2\Omega_{\varphi}(f)=1.67\times 10^{-5} \tilde{\beta}^{-2}\left(\frac{\kappa_{\varphi} \alpha}{1+\alpha}\right)^2
\left(\frac{100}{g_{*}}\right)^{1/3}\left(\frac{0.11v^3_{w}}{0.42+v^2_w }\right)S_{\varphi}(f),
 \label{omephi}
\end{align}
 where the spectral shape of the peak frequency $f_{\varphi}$ is
 \begin{align}
  \label{Sphi}
S_{\varphi}(f) & =\frac{3.8(f/f_{\varphi})^{2.8}}{1+2.8(f/f_{\varphi})^{3.8}}
\end{align}
with the peak frequency 
 \begin{align}
 \label{fpeakphi}
  f_{\varphi} & = 16.5 \times 10^{-6} \tilde{\beta} \left(\frac{0.62}{1.8-0.1v_w+v^2_w }\right)\left( \frac{T_{\mathrm{t}}}{100~\mathrm{GeV}}\right)\left( \frac{g_{*}}{100}\right)^{1/6}~\mathrm{Hz}.
\end{align}
%===============
  \item Sound wave contribution $\Omega_{\mathrm{sw}}$:
 \begin{align}
h^2\Omega_{\mathrm{sw}}(f)=2.65\times 10^{-6} \tilde{\beta}^{-1}\left(\frac{\kappa_v \alpha}{1+\alpha}\right)^2\left(\frac{100}{g_{*}}\right)^{1/3}v_{w}S_{\mathrm{sw}}(f),
 \label{omesw}
\end{align}
 where the spectral shape of the peak frequency $f_{\mathrm{sw}}$ is
 \begin{align}
  \label{Ssw}
S_{\mathrm{sw}}(f) & =(f/f_{\mathrm{sw}})^3\left(\frac{7}{4+3(f/f_{\mathrm{sw}})^2}\right)^{7/2}
\end{align}
with the peak frequency
 \begin{align}
 \label{fpeak}
  f_{\mathrm{sw}} & = 1.9 \times 10^{-5} v_w ^{-1} \tilde{\beta}\left( \frac{T_{\mathrm{t}}}{100~\mathrm{GeV}}\right)\left( \frac{g_{*}}{100}\right)^{1/6}~\mathrm{Hz}.
\end{align}
%===============
  \item MHD turbulence contribution $\Omega_{\mathrm{turb}}$:
 \begin{align}
h^2\Omega_{\mathrm{turb}}(f)=3.35\times 10^{-4} \tilde{\beta}^{-1}\left(\frac{\kappa_{\mathrm{turb}} \alpha}{1+\alpha}\right)^{\frac{3}{2}}\left(\frac{100}{g_{*}}\right)^{1/3}v_{w}S_{\mathrm{turb}}(f),
 \label{ometurb}
\end{align}
 where the spectral shape of the peak frequency $f_{\mathrm{turb}}$ is
 \begin{align}
  \label{Sturb}
S_{\mathrm{turb}}(f) & =\frac{(f/f_{\mathrm{turb}})^3}{[1+(f/f_{\mathrm{turb}})]^{\frac{11}{3}}(1+8\pi f/h_{\mathrm{t}})}
\end{align}
with the peak frequency
 \begin{align}
 \label{fpeakturb}
  f_{\mathrm{turb}} & = 2.7 \times 10^{-5} v_w ^{-1} \tilde{\beta}\left( \frac{T_{\mathrm{t}}}{100~\mathrm{GeV}}\right)\left( \frac{g_{*}}{100}\right)^{1/6}~\mathrm{Hz},
\end{align}
and  \begin{align}
  \label{h}
h_{\mathrm{t}}& = 16.5 \times 10^{-6} \left( \frac{T_{\mathrm{t}}}{100~\mathrm{GeV}}\right)\left( \frac{g_{*}}{100}\right)^{1/6}~\mathrm{Hz},
\end{align}
which is the value (redshifted to today)  of the inverse Hubble time at the GW production.
\end{itemize}

Bubbles produced by quantum tunneling grow with velocity $v_w$.
It is even possible for $v_w$ to approach 
continuously  to the speed of light
(runaway configuration)
 \cite{Bodeker:2009qy,Bodeker:2017cim}.
In a no-runaway  case, the bubble wall velocity $v_w$
terminates at a certain  velocity $< 1$.
The criterion for runaway bubbles is the value
of $\alpha$ compared with  $\alpha_{\infty}$
(the minimum value of $\alpha$ for  runaway bubbles):
\begin{itemize}
  \item $\alpha_{\infty}>\alpha$: No runaway bubbles ($h^2\Omega_{\mathrm{GW}}\simeq h^2\Omega_{\mathrm{sw}}+h^2\Omega_{\mathrm{turb}}$)
  \item $\alpha_{\infty}<\alpha$: Runaway bubbles  ($h^2\Omega_{\mathrm{GW}}\simeq h^2\Omega_{\mathrm{\varphi}}+h^2\Omega_{\mathrm{sw}}+h^2\Omega_{\mathrm{turb}}$),
\end{itemize}
where  $\alpha_{\infty}$ is given by  \cite{Espinosa:2010hh,Caprini:2015zlo}
\begin{align}
\alpha_{\infty}& \simeq \frac{30}{24\pi^2}\frac{\sum_a c_a\Delta m^2 _a(\varphi)}{g_* T^2_{\mathrm{t}}}.
\end{align}
Here $c_a$ is the degree of freedom of the particle $a$
(which should be multiplied with $1/2$ in the fermionic case  in addition),
and $\Delta m^2 _a(\varphi)$ is the difference of its field-dependent 
squared masses in  two phases.
For our model with $g_*=115.75$ we use
\begin{align}
\alpha_{\infty}& \simeq 1.09\times10^{-3}\left[n_f\left(\frac{2M(\left<\varphi_i\right>)}{T_{\mathrm{t}}}\right)^2+3\lambda_{S}\left(\frac{\left<S\right>}{T_{\mathrm{t}}}\right)^2\right].
\end{align}
Here we have used the relation
 $m^2_{\sigma}\simeq (2M)^2$, where 
 the constituent mass $M$ is given in Eq. (\ref{M}).
 This relation is approximately satisfied, because we have neglected the contribution
 from the Yukawa coupling $y$ (which is very small for our benchmark parameters).
We have computed $\alpha_{\infty}$  for the 
benchmark points and found
\begin{align}
\label{ainfty}
\alpha_{\infty}= \left\{\begin{array}{cc}\text{Case A :}~0.116~~~~ & \text{Case C :} ~0.125~~\\\text{Case B :} ~0.092~~~~&\text{Case D :}~ 0.095~~\end{array}\right..
\end{align}
Comparing $\alpha$
 given in Table \ref{GWparas} with  $\alpha_{\infty} $
for each benchmark point we see  that the bubbles  for all cases
 run away. With $\alpha_{\infty}$ given above we can then
compute the fraction $\kappa$ of the latent heat converted 
to the relevant contribution to the GW spectrum \cite{Caprini:2015zlo}: 
\begin{align}
\label{runwayintf}
\kappa_{\varphi}\equiv 1- \frac{\alpha_{\infty}}{\alpha},~~~~\kappa_v\equiv \frac{\alpha_{\infty}}{\alpha}\kappa_{\infty},~~~~\kappa_{\mathrm{turb}}=\epsilon\kappa_v,~~~~\kappa_{\infty}\equiv \frac{\alpha_{\infty}}{0.73+0.083\sqrt{\alpha_{\infty}}+\alpha_{\infty}},
\end{align}
where for all the benchmark cases
 (being all  runaway)
we have assumed that  the wall velocity $v_w$ is close to the speed of light,
and $\epsilon=0.05$
\cite{Caprini:2015zlo} for the MHD turbulence.
With Eqs.~(\ref{omephi})--(\ref{h}), (\ref{ainfty}), and (\ref{runwayintf})
we are now in position to compute
the GW signal for the benchmark cases.\footnote{The resent papar \cite{Jinno:2017ixd} is
 certainly relevant to us, but the paper has 
 appeared after we completed our calculations. }

\begin{figure}[t]
\begin{minipage}{0.49\hsize}
\begin{center}
\includegraphics[width=3.2in]{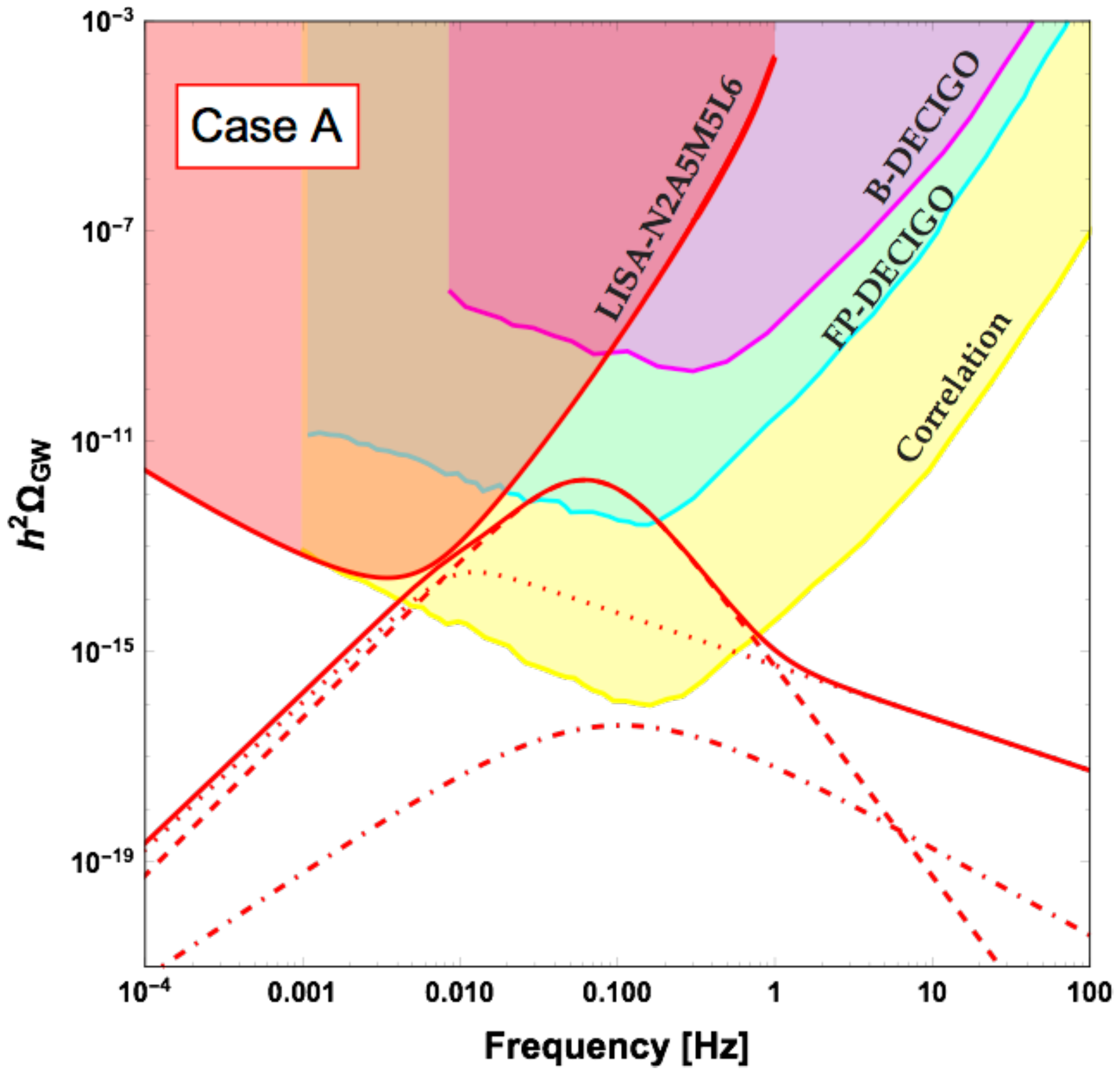}
\end{center}
\end{minipage}
\begin{minipage}{0.49\hsize}
\begin{center}
\includegraphics[width=3.2in]{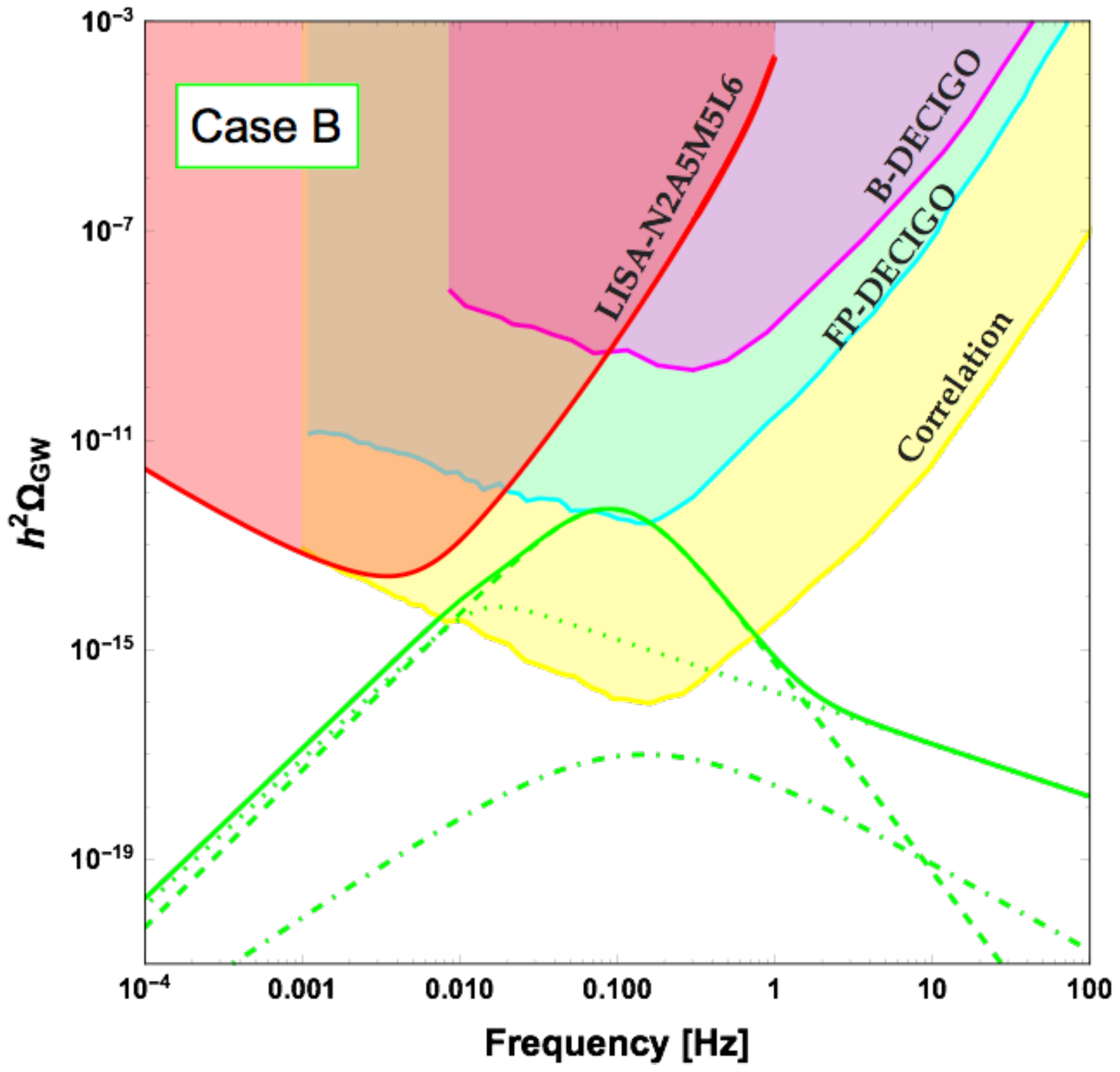}
\end{center}
\end{minipage}
\begin{minipage}{0.49\hsize}
\begin{center}
\vspace{+2mm}
\includegraphics[width=3.2in]{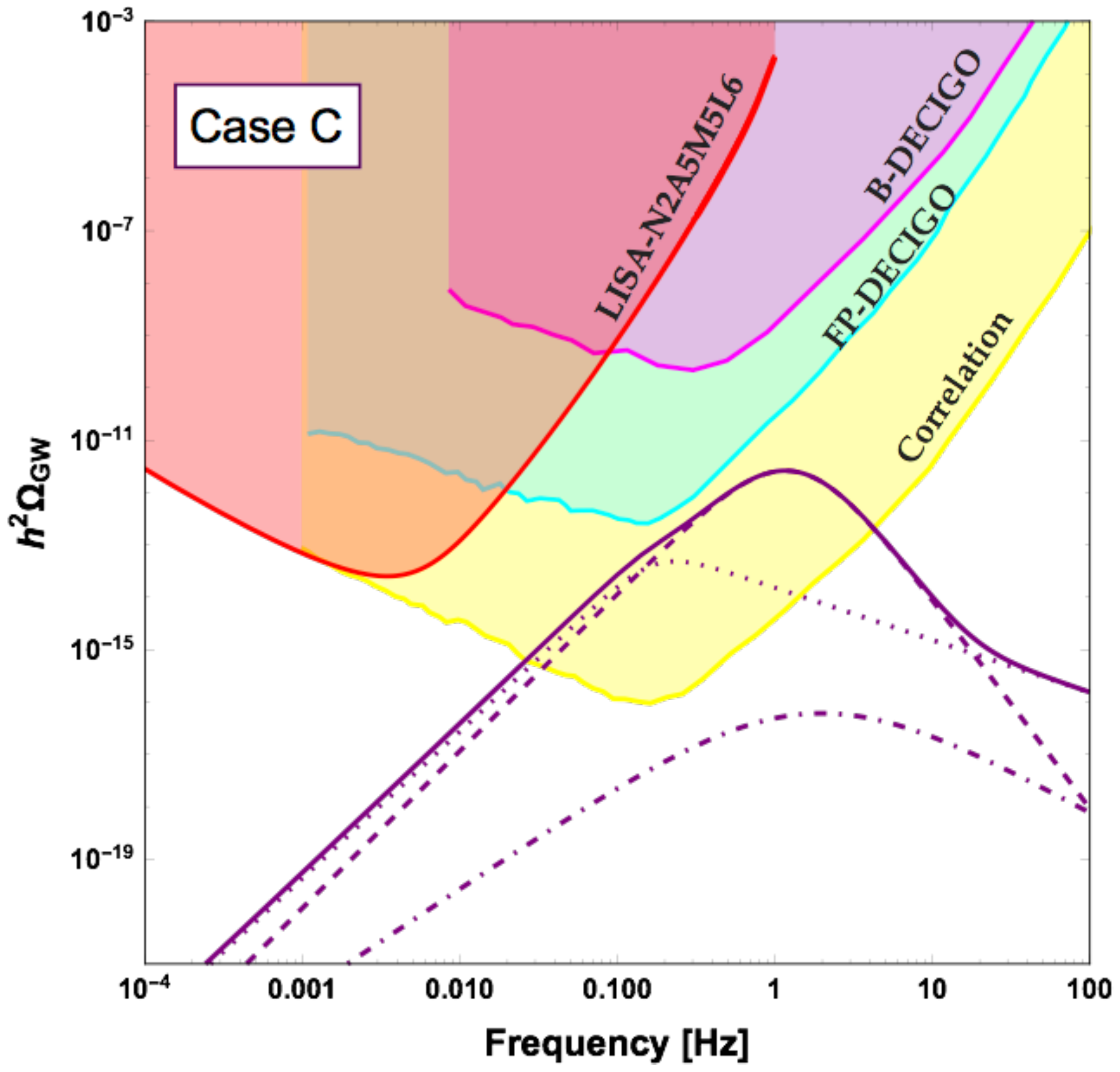}
\end{center}
\end{minipage}
\begin{minipage}{0.49\hsize}
\begin{center}
\vspace{+2mm}
\includegraphics[width=3.2in]{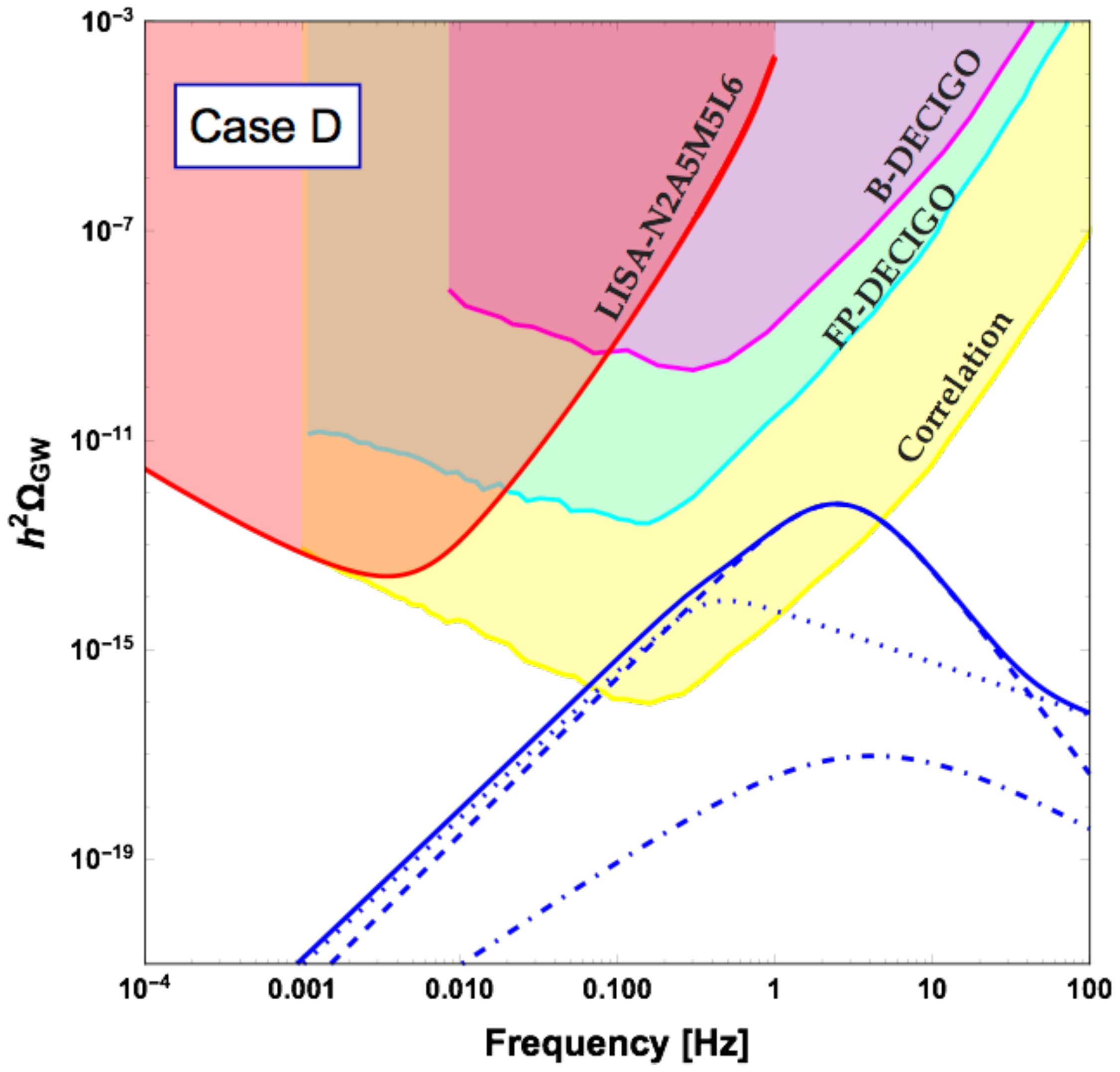}
\end{center}
\end{minipage}
\vspace{+2mm}
\caption{The GW spectrum with $v_w=1$ for Case A (top left), B (top right), C (bottom left), and D (bottom right).  The total GW spectrum (solid lines) is the sum of
 the sound wave (dashed lines), scalar (dotted lines), and MHD turbulence (dashed-dotted lines) contributions. The colored regions are observable  regions of LISA (LISA-N2A5M5L6 \cite{Caprini:2015zlo}) and DECIGO (B-DECIGO, FP-DECIGO, and Correlation \cite{Seto:2001qf,Kawamura:2006up,Kawamura:2011zz}).}
\label{sigABCD}
\end{figure}

\begin{figure}[t]
\begin{center}
\includegraphics[width=5.4in]{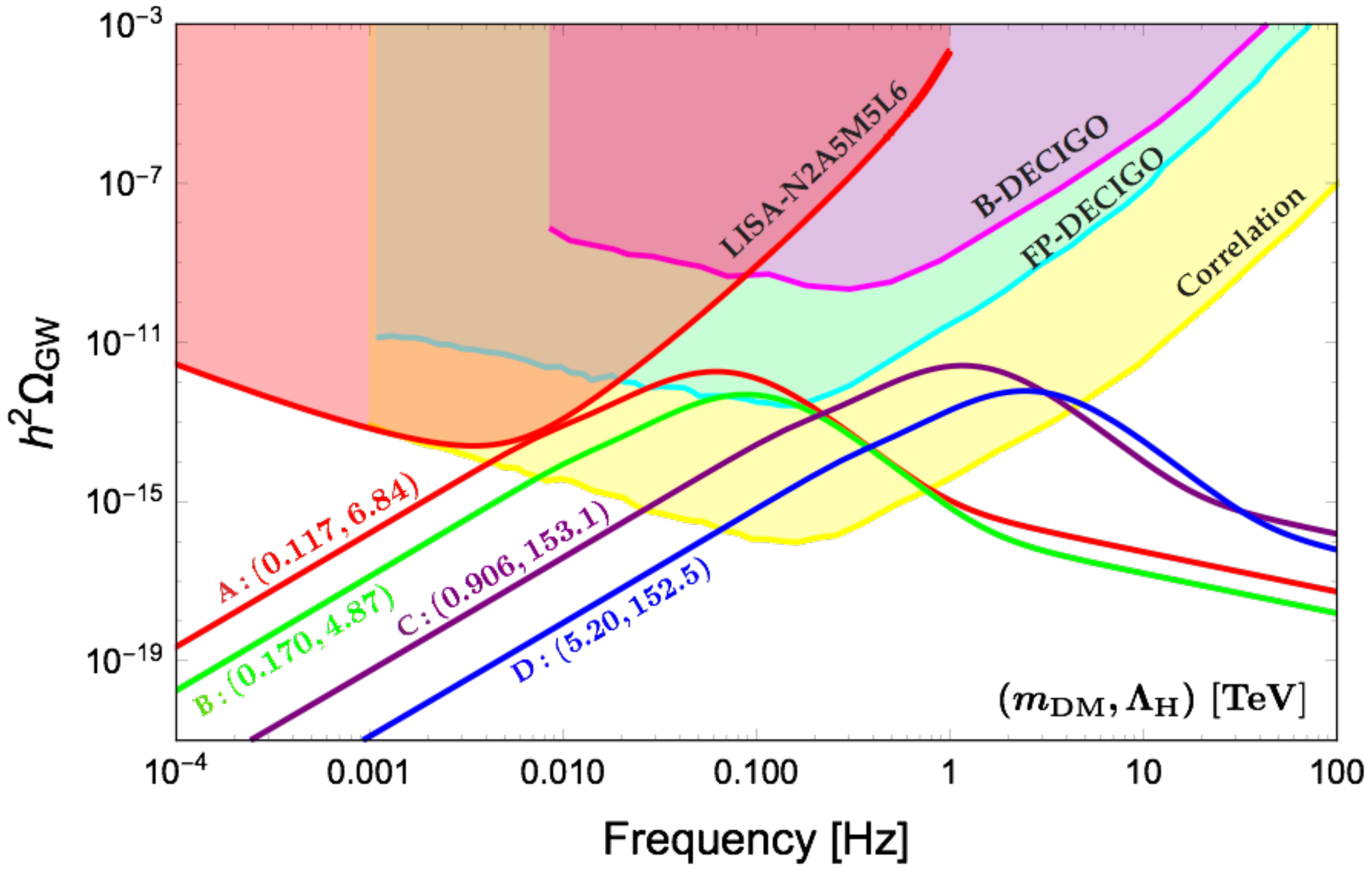}
\vspace{-2mm}
\caption{The GW spectrum with $v_w=1$ for Case A (red), B (green), C (purple), and D (blue). The numbers in the parentheses are
$m_{\mathrm{DM}}$ and $\Lambda_{\mathrm{H}}$ in units of TeV
(Table \ref{CaseABCD}).
The colored regions are observable regions of LISA (LISA-N2A5M5L6 \cite{Caprini:2015zlo}) and DECIGO (B-DECIGO, FP-DECIGO, and Correlation \cite{Seto:2001qf,Kawamura:2006up,Kawamura:2011zz}).}
\label{ABCDSigrun}
\end{center}
\end{figure}

In Fig.~\ref{sigABCD} we present our results.
For each benchmark case (A--D) we show the GW spectrum with $v_w = 1$, where
the total GW signal, 
sound wave, scalar, and  MHD turbulence contributions 
are denoted by the solid, dashed, dotted, and dashed-dotted lines, respectively.
The colored regions show observable regions  of different configurations
of LISA 
\cite{Caprini:2015zlo,Audley:2017drz}  and DECIGO \cite{Seto:2001qf,Kawamura:2006up,Kawamura:2011zz}.
The label of ``LISA-N2A5M5L6" corresponds to the configuration of LISA provided in Table 1 in  \cite{Caprini:2015zlo},
while the labels ``B-DECIGO," ``FP-DECIGO," and ``Correlation" are DECIGO designs \cite{Seto:2001qf,Kawamura:2006up,Kawamura:2011zz}.
As we can see from Fig.~\ref{sigABCD}, the sound wave contribution is dominant
for all the cases, while the MHD turbulence contribution is negligibly
small, so that the peak frequency of the GW spectrum is basically that of  
the sound wave contribution. 
The contribution MHD turbulence is small because
$\epsilon$ (the fraction of turbulent bulk motion) is set to $0.05$ \cite{Caprini:2015zlo}.
The scalar contribution becomes non-negligible
at higher frequencies and consequently changes the slope for this region
of frequency. But since it depends on $\tilde{\beta}^{-2}$ [see Eq.~(\ref{omephi})],
the sound wave contribution being proportional
to $\tilde{\beta}^{-1}$ is larger for smaller
$\tilde{\beta}$.
The peak frequencies of Cases A and B are $\sim 0.1$ Hz, while those of C and D are a few hertz. The main reason for this difference 
is the different transition temperature $T_{\mathrm{t}}$ (see Table II),
which once again results from the difference of 
$\lambda_{HS}$ (see Table I).
Consequently, the GW signal is difficult to observe at LISA \cite{Caprini:2015zlo,Audley:2017drz}.
The peak values of the GW spectrum are $10^{-12}$ for A and C, while those
for B and D are $10^{-13}$. Therefore, DECIGO sensitivities 
\cite{Seto:2001qf,Kawamura:2006up,Kawamura:2011zz} may be sufficient to
observe the signal.
Finally we summarize the results for Cases A--D in Fig.~\ref{ABCDSigrun} with the DM mass $m_{\mathrm{DM}}$ and the hidden QCD scale $\Lambda_{\mathrm{H}}$.
If Cases A and B, and also C and D, could be experimentally distinguished,
we could obtain information
about the magnitude of the explicit chiral symmetry breaking in the hidden sector.

%=============   \ref{sigABCD}   ==============%
%=================================%

%=============== Fig  ABCDSigrun   =============================
%=============== Fig HQCDgwASigrun   =============================

%=======================================================================%
%=======================================================================%
\section{Summary and Conclusion}\label{Summary and Conclusion}
Mass can be created by nonperturbative effects 
in non-Abelian gauge theories from nothing. 
By ``from nothing'' we mean that the theory has no dimensional
parameter and  hence is scale invariant at the classical level.
Scale invariance is broken explicitly by a scale anomaly
and at the same time dynamically by the nonperturbative effects.
Dynamical breaking of scale invariance
can be used to explain the origin of the Higgs mass as well as of the DM mass \cite{Hur:2007uz,Hur:2011sv,Heikinheimo:2013fta,Holthausen:2013ota,Kubo:2015cna,Ametani:2015jla,Kubo:2014ida,Hatanaka:2016rek,Kubo:2015joa,Ishida:2017ehu,Haba:2017wwn}.

Needless to say, dynamical breaking of scale invariance
is associated with a phase transition at finite temperature
\cite{Holthausen:2013ota,Kubo:2015joa,Tsumura:2017knk}.
If the phase transition is of first  order and strong enough
in the early Universe, it can produce GW which might be observed today 
as a  GW background \cite{Witten:1984rs}.

In this paper we have expanded our analysis of a particular scale invariant
extension of the SM to include the aspect of 
the GW background predicted by the model.
The model contains a strongly interacting hidden sector,
described by a non-Abelian gauge theory,
in which a mass scale in the TeV region is generated  through the chiral symmetry breaking
in the hidden sector.
The corresponding (pseudo) NG bosons are a realistic candidate for DM, since
their mass is finite because the chiral symmetry is also explicitly broken
by a Yukawa coupling between  the hidden sector fermions and  a SM singlet
real scalar field $S$.
The scalar field $S$ plays the role of a mediator that transfers the robust 
energy scale from the hidden sector to the SM sector
via a Higgs portal coupling.

As in \cite{Holthausen:2013ota,Ametani:2015jla} we have used the NJL method to  effectively treat 
the D$\chi$SB.
Integrating out  the hidden sector fermions in the NJL model
yields an effective potential for the chiral condensate  
at zero and finite temperature. In the mean field approximation we can identify
the chiral condensate with $\sigma$ and the NG bosons with $\phi_a$ 
(which are DM). We have restricted ourselves to $n_c = n_f = 3$ for the hidden sector
QCD, because we can simply scale up the parameters 
of the NJL model for the real QCD, 
such that the hidden sector NJL model has the same number of independent
parameters as that of the hidden sector QCD.

As it is known, the nature of the chiral phase transition changes depending
on the strength of the explicit chiral symmetry braking.
For the hidden sector QCD it means, on one hand, that the Yukawa coupling constant 
$y$ should be sufficiently small to obtain a strong first-order 
chiral phase transition. On the other hand, a small $y$
 implies two-stage
phase transitions; the chiral phase transition at $T \gsim \mathcal{O}(1)$ TeV and
the EW phase transition at $T\sim \mathcal{O}(100)$ GeV.
That is, two phase transitions can be clearly distinguished.

Using the technique in the literature (see 
\cite{Caprini:2015zlo} and references therein) within the framework
of the NJL model in the mean field approximation, we have analyzed
the GW background produced by the chiral phase transition in the hidden sector
of the model. In particular, depending on
the value of $y$ and of the Higgs portal coupling $\lambda_{HS}$, 
we have chosen four benchmark points
in the parameter space.
These points are representative points characterized by the magnitude of the 
explicit chiral symmetry breaking
and the hidden sector scale $\Lambda_{\mathrm{H}}$. 
We have found for these points that the peaks 
of the GW signal appear at frequencies $\mathcal{O}(0.01 - 1)$ Hz.
Unfortunately, these frequencies are slightly too high, so that 
it will be difficult  for them to be observed at LISA 
\cite{Caprini:2015zlo,Audley:2017drz}.
But their strength seems to be sufficiently large 
for observations  at  DECIGO 
\cite{Seto:2001qf,Kawamura:2006up,Kawamura:2011zz} 
which will cover a higher frequency region.
We emphasize that observation of a GW background signal at  
frequencies $0.1 \sim \mbox{few}$ hertz
with $h^2\Omega_{\mathrm{GW}}  \gsim 10^{-13}$
may be a strong indication 
for  strongly interacting hidden sector models.

Finally we should admit that
our results have been  obtained by using the NJL model,
which is supposed to  serve as an effective theory 
of the hidden sector QCD. A fair question is about the
systematic uncertainties present in this approach.
At the moment we can say only that the NJL model for the
real hadrons can reproduce their basic quantities with an uncertainty of
$\mathcal{O}(10-20)$\% \cite{Hatsuda:1994pi}. Therefore, to make more precise predictions
it is certainly inevitable to use a more reliable method such 
as lattice gauge theory.

%=======================================================================%
\section*{Acknowledgments}
We thank Thomas Konstandin for useful discussions.
The work of M.~A. is supported in part by the Japan Society for the
Promotion of Sciences Grant-in-Aid for Scientific Research (Grants
No. 16H00864 and No. 17K05412). 
J.~K.~is partially supported by the Grant-in-Aid for Scientific Research (C) from the Japan Society for Promotion of Science (Grant No. 16K05315).

%=======================================================================%
%
%  Appendix  
%
%=======================================================================%
%=======================================================================%
\begin{appendix}
%=======================================================================%

%=======================================================================%
 \section{Thermal Function for Field Renormalization Constant}
The field renormalization can be computed as
\begin{align}
\nonumber
   Z^{-1} _{\sigma}(S,\sigma)  & = -\left( 1-\frac{G_D}{4G^2}\sigma \right)^2 3 n_c \left. \frac{d}{dp^2}I_{\varphi^2}(p^2,M;\Lambda_{\mathrm{H}})\right|_{p^2=0},
\end{align}
where the loop function $I_{\varphi^2}(p^2,M)$ is given in Eq.~(\ref{propagator for sigma}) and its derivative can be written as
 \begin{align}
 \nonumber
&  \left.\frac{d}{dp^2}I_{\varphi^2}(p^2,M)\right|_{p^2=0}\\
\nonumber
  & = -4\int \frac{d^4 k}{i(2\pi)^4}\frac{1}{(k^2-M^2)^2}+ 4\int \frac{d^4 k}{i(2\pi)^4}\frac{2M^2}{(k^2-M^2)^3}
  \equiv -4 I_A(M)+4I_B(M),
 \end{align}
 where we defined two terms as $I_{A}$ and $I_B$.
 Using the standard calculation method at finite temperature,
 they can be computed as
 \begin{align}
I_{A}& =\frac{T}{2\pi i} \oint_{C} \frac{d^3 k}{(2\pi)^3}\frac{1}{(k_0 ^2-\omega^2)^2}\frac{1}{2}\beta\tanh \left(\frac{1}{2}\beta k_0 \right)= A^{0}_F(M;\Lambda_{\mathrm{H}})+A_F(u^2),\\
I_{B}& = \frac{T}{2\pi i} \oint_{C} \frac{d^3 k}{(2\pi)^3}\frac{2M^2}{(k_0 ^2-\omega^2)^3}\frac{1}{2}\beta\tanh \left(\frac{1}{2}\beta k_0 \right)=B^{0}_F(M;\Lambda_{\mathrm{H}})+B_F(u^2),
 \end{align}
 where $\beta=1/T$, $k_0=i\omega_n=i\pi(2n+1) T$,
 $u=M/T$, and
 the function $\frac{1}{2}\beta \tanh(\frac{1}{2}\beta k_0)$ has a pole at $k_0$.
 The zero-temperature components with four-dimensional cutoff are
 \begin{align}
 A^{0}_F(M;\Lambda_{\mathrm{H}})&= \int_{\Lambda_{\mathrm{H}}}\frac{d^4 k_E}{(2\pi)^4}\frac{1}{(k_E ^2+M^2)^2} = \frac{1}{(4\pi)^2}\frac{1}{2}\left[\ln \left(1+\frac{\Lambda^2 _{\mathrm{H}}}{M^2}\right)-\frac{\Lambda^2_{\mathrm{H}}}{\Lambda^2_{\mathrm{H}}+M^2} \right],\\
 B^{0}_F(M;\Lambda_{\mathrm{H}})&= -\int_{\Lambda_{\mathrm{H}}}\frac{d^4 k_E}{(2\pi)^4}\frac{2M^2}{(k_E ^2+M^2)^3}=  -\frac{1}{(4\pi)^2}\frac{\Lambda^4_{\mathrm{H}}}{2(\Lambda^2_{\mathrm{H}}+M^2)^2},
   \end{align}
and those thermal effect functions can be written as
 \begin{align}
 \nonumber
 A_F(u^2) & =\int^{i\infty+\epsilon} _{-i\infty+\epsilon}\frac{dk_0}{2\pi i}\int \frac{d^3k}{(2\pi)^3}\frac{2}{(k_0-\omega)^2(k_0+\omega)^2}\frac{1}{e^{\beta k_0}+1}\\
 \nonumber
 & =-\frac{1}{4\pi^2}\int_0 ^{\infty} dx \frac{x^2}{(\sqrt{x^2+u^2})^3}\frac{1}{1+e^{\sqrt{x^2+u^2}}}\\
 \label{AFT}
 &~~~~-\frac{1}{8\pi^2}\int_0 ^{\infty} dx \frac{x^2}{(\sqrt{x^2+u^2})^2}\frac{1}{1+\cosh\sqrt{x^2+u^2}},\\
 \nonumber
 B_F(u^2) & = \int^{i\infty+\epsilon} _{-i\infty+\epsilon}\frac{dk_0}{2\pi i}\int \frac{d^3k}{(2\pi)^3}\frac{4M^2}{(k_0-\omega)^3(k_0+\omega)^3}\frac{1}{e^{\beta k_0}+1}\\
 \nonumber
& =\frac{2u^2}{\pi^2}\left[ 3\int^{\infty}_0 d x \frac{ x^2}{(\sqrt{x^2+u^2})^5}\frac{1}{1+e^{\sqrt{x^2+u^2}}}\right. \\
 \nonumber
&\left.~~~~+3\int^{\infty}_0  dx\frac{x^2}{(\sqrt{x^2+u^2})^4} \frac{1}{1+\cosh \sqrt{x^2+u^2}}\right.\\
\label{BFT}
&\left.~~~~+\int^{\infty}_0  d x\frac{x^2}{(\sqrt{x^2+u^2})^3} \frac{1}{1+\cos\sqrt{x^2+u^2}}\tanh \left(\frac{1}{2}\sqrt{x^2+u^2}\right)\right].
 \end{align}
 In this work we fitted each thermal function using the following fitting functions,
  \begin{align}
 A_F(u^2)& =\frac{1}{8\pi^2}\ln u+ e^{-u}\sum^{40} _{n=0} a_n u^{n},\\
 B_F(u^2) &=e^{-u}\sum^{40} _{n=0} b_n u^{n}.
 \end{align}

%=======================================================================%
\end{appendix}

%=======================================================================%
%=======================================================================%

%=======================================================================%
\end{document}